\documentclass{aa}  

\usepackage{graphicx}
\usepackage{txfonts}
\usepackage{ulem}
\usepackage{graphicx}
\usepackage{tabularx}
\usepackage{txfonts}
\usepackage{longtable}
\usepackage{natbib}
\usepackage{hyperref}
\usepackage{verbatim}
\usepackage{url}
\usepackage{float,color}
\usepackage{subcaption}
\usepackage[figuresright]{rotating}
\hypersetup{
    colorlinks=true,
    linkcolor=blue,
    filecolor=magenta,      
    urlcolor=cyan,
    citecolor=blue
}

\bibpunct{(}{)}{;}{a}{}{,}
\newcommand{\uproman}[1]{\uppercase\expandafter{\romannumeral#1}}
\newcommand{\ur}[1]{\uproman{#1}}

\newcommand{\ha}{\ensuremath{{\rm H_{\alpha}}}}
\newcommand{\hb}{\ensuremath{{\rm H_{\beta}}}}

\newcommand{\teff}{\rm T_{\rm eff}}
\newcommand{\logg}{\log g}

\newcommand{\ltross}{\rm \log \tau_{ross}}

\newcommand{\opd}{\log \tau_{\rm 5000}}

\newcommand{\dis}{\texttt{DISPATCH}\ }

\newif\ifhighlight
\highlightfalse 

\begin{document}

\title{M3DIS - A grid of 3D radiation-hydrodynamics stellar atmosphere models for stellar surveys}

\subtitle{\uproman{1}. Procedure, Validation \& The Sun}

\author{Philipp Eitner \inst{1,2}, Maria Bergemann \inst{1,3}, Richard Hoppe \inst{1,2},  $\rm \AA$ke Nordlund \inst{3}, Bertrand Plez \inst{4}, Jonas Klevas \inst{1}}

\institute{
Max Planck Institute for Astronomy, 69117 Heidelberg, Germany
\label{1}
\and
Heidelberg University, Grabengasse 1, 69117 Heidelberg, Germany
\label{2}
\and
Niels Bohr International Academy, Niels Bohr Institute, University of Copenhagen, Blegdamsvej 17, DK-2100 Copenhagen, Denmark
\label{3}
\and
Laboratoire Univers et Particules de Montpellier, University of Montpellier, CNRS, Montpellier, France
\label{4}}
\date{}

 
  \abstract
   {Large-scale stellar surveys, such as SDSS-V, 4MOST, WEAVE, and PLATO, require accurate atmospheric models and synthetic spectra of stars for accurate analyses of fundamental stellar parameters and chemical abundances.}
   {The primary goal of our work is to develop a new approach to solve radiation-hydrodynamics (RHD) and generate model stellar spectra in a self-consistent and highly efficient framework.}
   {We build upon the Copenhagen legacy RHD code, the \texttt{MULTI3D} non-local thermodynamic equilibrium (NLTE) code, and the \texttt{DISPATCH} high-performance framework. The new approach allows us to calculate 3D RHD models of stellar atmospheres on timescales of a few thousand CPU hours and to perform subsequent spectrum synthesis in local thermodynamic equilibrium (LTE) or NLTE for the desired physical conditions within the parameter space of FGK-type stars.}
   {We compare the 3D RHD solar model with other available models and validate its performance against solar observations, including the centre-to-limb variation of intensities and key solar diagnostic lines of H and Fe. We show that the performance of the new code allows to overcome the main bottleneck in 3D NLTE spectroscopy and enables calculations of multi-dimensional grids of synthetic stellar observables for comparison with modern astronomical observations.}
   {}

\titlerunning{3DNLTE-Grid}
\authorrunning{Eitner et al.}
\maketitle
\section{Introduction} \label{sec:introduction}
Energy transfer below the optical surface of atmospheres of late A- and FGKM-type stars is dominated by convection \citep{Chandrasekhar1939, Schwarzschild1958}. Convection defines the inner properties of stars but it also has a direct coupling to the physical structure and dynamical evolution of their atmospheric layers, thereby having a major impact on observable electromagnetic spectra, their photometric magnitudes, and light curves, thereby setting a fundamental scale for the characterisation of the surface and interior structure of stars via spectroscopy, photometry, and asteroseismology techniques \citep[e.g.][]{Nordlund2009, Ludwig2016, Bonifacio2018, Mosumgaard2020, Lind2024}.

Most studies of stellar atmospheres have been traditionally carried out using one-dimensional (1D) models in hydrostatic equilibrium (HE) and local thermodynamic equilibrium (LTE) \citep[e.g.][]{Kurucz1979, Grupp2004, Gustafsson2008}. These strong approximations allow for rather simple and computationally affordable models, yet at the expense of physical accuracy. Convection is an inherently three-dimensional (3D) phenomenon and it is therefore highly approximate in such models, as convective flux is represented by parameters based on the mixing-length theory or its variants \citep{Boehm1958,Canuto1991} and motions on small and large scales by ad hoc parameters of micro and macroturbulence \citep{Gray1992}, respectively. Codes for 3D radiation-hydrodynamic (RHD) calculations exist, for example \texttt{STAGGER} \citep{Nordlund1982,Nordlund1995,Collet2011}, \texttt{MURaM} \citep{Vogler2004,Vogler2005} and \texttt{CO$^{5}$BOLD} \citep{Freytag2012}. However, 1D HE models have a tremendous computational advantage over 3D RHD based models. In comparison with the former models that can be computed on timescales of approximately a few CPU minutes on a laptop, the latter need of the order $1\,000$ (low spatial resolution) to $10\,000$ (high resolution) CPU hours to achieve a robust relaxed local model of the solar outer layers, whereby realistic local models of metal-poor red giants may require up to a few $100\,000$ CPU hours (priv. comm. H.-G. Ludwig, C. Lagae) or more, around $1\,000\,000$ CPU hours, for global models of asymptotic giant branch (AGB) stars or red supergiants (priv. comm. B. Freytag, S. H\"{o}fner). Furthermore, the total number of CPU hours depends highly on different properties, such as the number and distribution of opacity bins, geometric extent of the box and step size, number of rays for the angular resolution of the radiation, desired precision of the output parameters (such as $\teff$) and the length of the time series.  In NLTE, also  other complexities arise, including the convergence criteria, completeness of atomic models, and energy discretisation of all radiative quantities in the equations (Bergemann \& Hoppe in prep). While the physical aspects of the problem can be addressed with codes like \texttt{MULTI3D} \citep{Leenaarts2009} and \texttt{NLTE3D} \citep{Sbordone2010}, only studies of the chemical composition of the Sun \citep[e.g.][]{Asplund2009, Caffau2011a, Bergemann2021} and of a handful individual stars  \citep[e.g.][]{Caffau2011b, Nordlander2017, Lagae2023} were carried out in 3D NLTE framework. Simply put, to compute a 3D NLTE grid of model spectra with similar properties to those that are presently employed in surveys, as for APOGEE \citep{Meszaros2012}, one would need approximately $10^{11}$ CPU hours, to be compared to 10$^5$ CPU hours for a 1D HE model grid (which is doable on a medium-size cluster within a month), which explains why 1D LTE models are still widely and exclusively used in the analysis of million star datasets from large-scale stellar surveys.

In this work, we attempt to overcome one of the main bottlenecks that has limited progress in this area. Specifically we focus on developing a set of codes, within the scope of the \dis high-performance computing framework \citep{Nordlund2018}, in order to enable fast and efficient 3D simulations of stellar atmospheres and subsequent physical modelling of high-resolution stellar spectra within the same consistent approach, flexible as to handling of chemical mixture, opacities, resolution, but also resulting synthetic observables as required by modern spectral analysis methods, both classic \citep{Schoenrich2014} and machine learning based \citep[e.g.][]{Gent2022}. Two grids of 3D RHD atmospheric models were computed by other groups \citep{Ludwig2009,Magic2013a}. However, these grids are not publicly available and thus cannot be readily used by the community for testing and scientific exploitation. Also, the grids were computed with given assumptions on microphysics. For example, the range of chemical composition in the CIFIST grid \citep{Ludwig2009} is limited to four metallicity values, the \texttt{STAGGER} grid to seven \citep{Magic2013a}. Our approach, as outlined in this paper, is flexible and potentially allows the development of a robust diagnostic approach for large-scale stellar analyses as required by next-generation spectroscopic surveys, particularly 4MOST and high-resolution spectroscopic survey of the Galactic disc \& bulge stars, which we are co-leading \citep{Bensby2019} and which provides a strong incentive for the effort presented here.

The paper is organised as follows. In Sect. \ref{sec:methods}, we present the basic equations used to solve the RHD problem, the microphysics, including the opacities and the binning scheme, and the initial conditions. We further summarise the approach to the calculation of detailed synthetic spectra. The resulting simulations of the solar sub-surface convection are described in Sect. \ref{sec:results}, along with a series of validation tests addressing the impact of numerical resolution, abundances, and opacities. Finally, we discuss some of the synthetic observables and their sensitivity to the properties of the models, and close with conclusions and outlook in Sect. \ref{sec:discussion}.

\section{Methods} \label{sec:methods}
\subsection{Radiation-hydrodynamics} \label{subsec:hydro}

The 3D model atmospheres in this work are calculated within the \dis high-performance simulation framework \citep{Nordlund2018}. \dis incorporates existing solvers for RHD problems and significantly increases their speed with its efficient task-based parallelisation scheme that provides vectorisation, local timestepping and unlimited scaling with OpenMP and MPI. 

The detailed physics and numerical methods of RHD problems have been described in many previous studies \citep[e.g.][]{Stein1998, Fromang2006,Freytag2012}. In the following, we will briefly summarise the main aspects that are relevant to this work. The set of equations that are solved to advance the simulation in time include the continuity, momentum, and energy conservation equations. The latter is coupled to radiation through the radiative heating term, which itself requires the solution of the radiative transfer (RT),
\begin{align}
    \partial_{t} \rho &= -\vec{\nabla} \left( \rho \vec{v} \right) \label{eq:continuity} \\
    \partial_{t} \left(\rho \vec{v} \right) &= -\vec{\nabla} \left( \rho \vec{v} \vec{v} \right) - \vec{\nabla} p + \rho \vec{g} \label{eq:euler}\\
    \partial_{t} e &= -\vec{\nabla} \left( (e+p)\vec{v} \right) + q_{rad} \ ,
    \label{eq:energy}
\end{align}
with density $\rm\rho$, velocity $\rm\vec{v}$, pressure $\rm p$, surface gravity $\vec{g}$, and energy per unit volume $\rm e$. In practice the fluxes occurring inside the right-hand-side divergence terms are evaluated with a Riemann (HLLC) solver, analogous to the one used in the \texttt{RAMSES} code \citep{Fromang2006} or, alternatively, with solvers analogous to the ones used in the \texttt{STAGGER} and \texttt{BIFROST} \citep{Gudiksen2011} codes. The radiative heating $\rm q_{rad}$ is defined as 
\begin{align}
    q_{rad} &= 4 \pi \rho \int \kappa_{\lambda} (J_{\lambda} - S_{\lambda}) d\lambda \ , \label{eq:heating}
\end{align}
that is the monochromatic absorption coefficient $\rm \kappa_{\lambda}$-weighted difference between the monochromatic mean intensity $\rm J_{\lambda}$ and the source function $\rm S_{\lambda}$, where $\lambda$ is the wavelength. Under the assumption of LTE, the source function is given by the Planck function $B_{\lambda}$, so $\rm S_{\lambda} = B_{\lambda}$. In the following, all opacities are given per unit mass of material. The models presented in this work do not include magnetic fields, but several MHD solvers are available in \texttt{DISPATCH}, such as \texttt{RAMSES/HLLD}, \texttt{STAGGER}, and \texttt{BIFROST}, with a separate module allowing for non-ideal MHD corrections \citep{Nordlund2018}.

The solution of the RHD equations is performed by the \dis implementation of the \texttt{RAMSES} Riemann-HLLC solver \citep{Fromang2006}, which itself is an extension of the MUSCL-Hancock scheme \citep[e.g.][]{vanLeer1977}. In short: An update of the hydrodynamic (HD) variables in this scheme consists of several steps. First, the cell-centred, volume-averaged primitive variables are advanced a half timestep in a non-conservative way using a Taylor expansion of the Euler system and slope-limited finite-difference spatial derivatives. The slopes are furthermore utilised to construct the cell-interface values of the primitive variables, which are then fed into the Riemann solver to compute the cell-interface fluxes. These fluxes are then used to advance the volume averaged quantities in time. 
A similar Riemann solver is also used in the \texttt{CO$^{5}$BOLD} code \cite[with multiple options for the time integration scheme,][]{Freytag2012}. Other RHD codes, such as \texttt{STAGGER}, \texttt{BIFROST} and \texttt{MURaM}, rely on explicit finite-difference schemes for spatial derivatives and a Runge-Kutta method \citep{Williamson1980} as temporal scheme. 
The main difference between a Riemann and a classical, non-Riemann solver is the treatment of the pressure term in Eq. \ref{eq:euler}. In the classical case the pressure gradient is included as a source term on the right-hand-side, which means sound waves are not propagated using advection and the system evolves entirely with the gas velocity. In Riemann solvers this is not the case. Here all characteristics of the system are advected consistently by including the pressure gradient in the to-be-advected system. In practice, this enables typical Riemann solvers to excel at capturing shocks and discontinuities and provide a better conservation of physical quantities. At the same time, Riemann solvers generally are less capable of handling smooth, semi-static regions. In these situations gravity and pressure forces cancel numerically down to machine-precision in classical solvers, because they are both treated identically as external sources. This is very different from the strategy followed by Riemann solvers, where pressure terms are included in the advection, while gravity remains an external force \citep[e.g.][]{Roe1986,LeVeque1997,Dullemond2011,Freytag2012}. Whether one or the other is more appropriate depends on their specific implementation and the problem at hand.

The main advantage of \dis over previous codes for simulating stellar spectra is its parallelisation scheme and numerical design. In short, one of the main reasons for the improved performance is the use of local timesteps. Previous hydrodynamical codes of any astrophysical application relied on synchronised advancement of time. The rate in stellar atmospheres is dominated by the radiative timestep. For a relaxed model atmosphere it is usually at the boundary of the optical surface, but for a relaxing model atmosphere it can be anywhere. The core and unique ability of \dis is to automatically devote additional resources to such areas to advance them at a lower timestep, while the rest of the model atmosphere can advance further in time using less resources. In practice, to do so, \dis splits the simulation domain in so-called 'patches' that contain the local values of either hydrodynamic or radiation variables based on their position in the atmosphere. Updating those variables by advancing them in time constitutes a 'task', which is executed by a task scheduler based on available computation resources and its status relative to its neighbouring patches. For example: A task that solves the RT under a certain angle is required to wait for its upwind neighbours to reach the same (or a later) time than the task itself, but not for tasks that are assigned to regions downwind. Likewise is the HD task required to wait for the RT task to provide the current heating rate. A task hence only needs to check the time of its immediate neighbours before it can be queued for the next update, which then can be executed once a CPU thread is available. \dis works best if there is an over-subscription of tasks compared to CPU threads, so that the CPU utilisation is near 100\% at all times, thereby allowing a highly efficient use of available computational resources.

The implementation in OpenMP and MPI allows this scheduler to utilise all available threads on multiple nodes effectively by relying on shared memory within MPI ranks and virtual tasks in the transition region between different ranks for streamlined inter-node communication when needed. This design also incorporates local timestepping naturally through the local time of each patch, which causes significant speed and stability improvements. The rate at which the stellar atmosphere is evolving in time is dominated by the radiative timestep, which in turn depends on the local hydrodynamic properties and radiation intensity through the radiative courant condition. Because of its design, \texttt{DISPATCH} is able to devote more computation resources to regions where the timestep is small, and hence minimise CPU idle times that would otherwise arise from non-synchronised timesteps. \citep{Nordlund2018}

\begin{figure}
    \includegraphics[width=\linewidth]{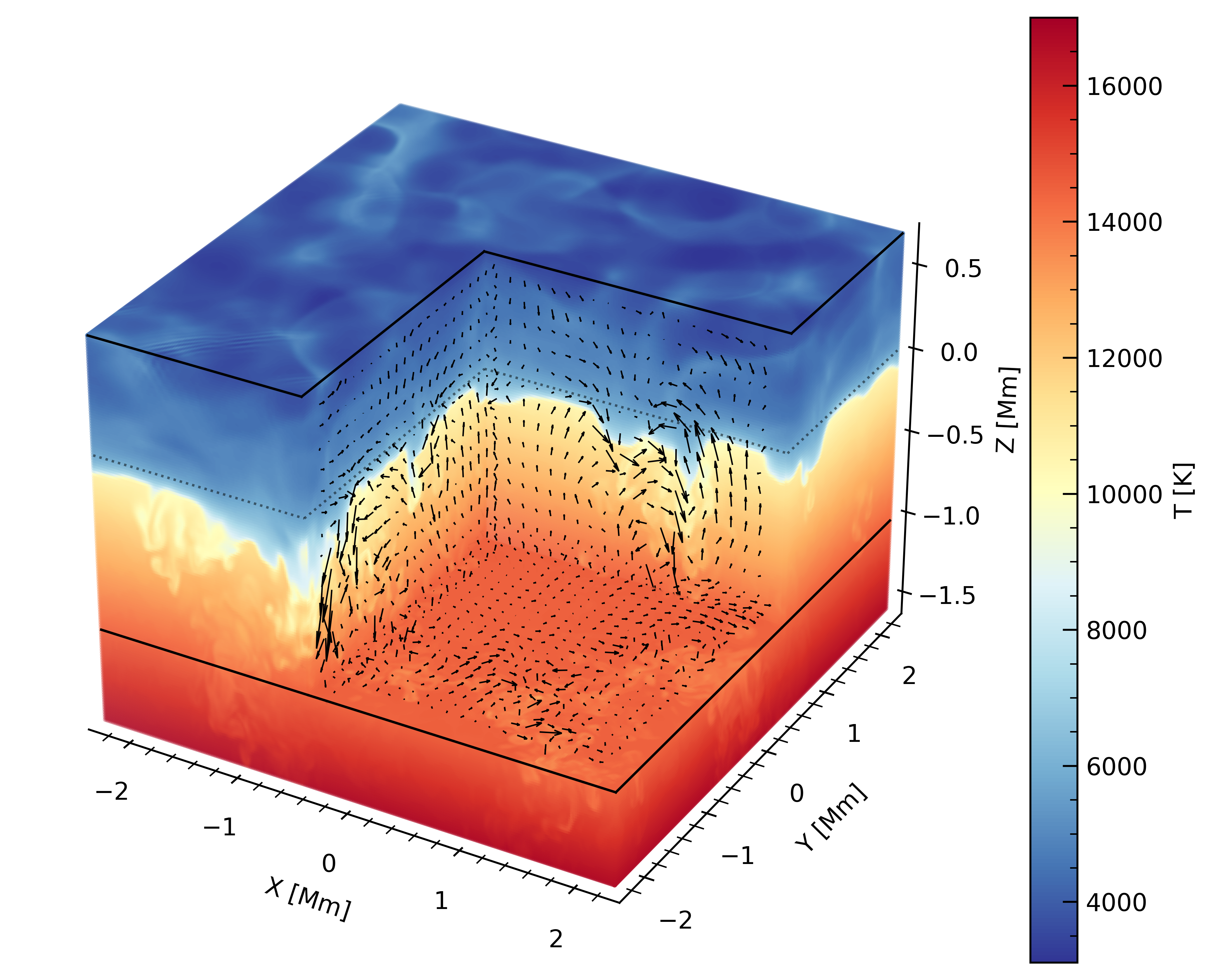}    
    \caption{Temperature structure for our reference solar 3D RHD model. Black arrows indicate the local velocity field. The corresponding animation of the time evolution of the RHD simulation is available as supplementary material to the online version of the paper.}
\label{fig:tcube}
\end{figure}

For illustration, in Fig. \ref{fig:tcube} we show the 3D distribution of temperature in the resulting solar simulation, which will be discussed in detail in Sect. \ref{sec:results}. The full 3D animation cube can also be retrieved as supplementary material to the online version of the article.
\subsection{Microphysics} \label{subsec:eos}
To close the system of RHD equations, an Equation of State (EoS) is required. In \dis the EoS is provided in tabular form. It provides the gas pressure ($\rm p_g$), electron density ($\rm N_e$), and temperature ($\rm T$) as a function of internal energy ($\rm e_{int}$) and density. The internal energy is hereby linked to Eq. \ref{eq:energy} though the subtraction of the kinetic energy, 
\begin{equation}
	e_{int} = e - \frac{1}{2} \ \rho \ \vec{v} \cdot \vec{v}.
\end{equation}
At a given density value, the internal energy depends on the temperature, chemical composition, as well as the ionisation and molecular equilibrium. To avoid potential inconsistencies with opacities, it is hence advisable to compute EoS and opacity tables simultaneously. Opacities themselves depend on the atomic and molecular level populations, which in LTE are functions of $\rm T$ and $\rm \rho$ only, and thus can be tabulated on the same grid as the EoS. Opacities are of utmost importance for the photospheric structure of stellar models, however their accurate treatment is one of the major computational complexities involved. A dense sampling of an extended wavelength grid from the UV to the IR is needed to capture most spectral features. Depending on the resolution of the $\rm T-\rho$ grid this requires significant computation time.
%
%
%
%

The opacities and EoS are computed using the \texttt{MARCS} \citep{Gustafsson2008} and \texttt{Turbospectrum} (TS) \citep{Plez2012, Gerber2023} codes.
They include the same routines for the calculation of the chemical equilibrium and of continuum opacities. For line opacities TS uses linelists while \texttt{MARCS} interpolates in opacity sampling tables generated for a set of temperatures and pressures on a fixed wavelength set.
They otherwise strictly use the same physical input data, for example partition functions, ionisation potentials, and dissociation energies. The version of the chemical equilibrium we use includes 92 atoms and their first two or three ions, as well as over 600 molecular species. 
A full description of the species included in the opacity package for continuum, atomic and molecular lines is given in \citet{Gustafsson2008}.
Line opacities are computed assuming LTE. Opacities are provided at the wavelengths used by \texttt{MARCS}, with a resolution of 
$\rm R =\lambda / \Delta \lambda  = 20,000$.

We sample a parameter space of $\rm 159 \times 159$ $\rm T-\rho$ points, equally spaced in natural logarithm, where $\rm T$ ranges between $\rm 1\,100\ K$ and $\rm 550\,000\ K$, and $\rm \rho$ between $\rm 10^{-15}\ g\times  cm^{-3}$ and $\rm 10^{-3}\ g\times cm^{-3}$. These limits are motivated by the typical temperature and density extend of solar metallicity FGK-type stars, however they are extended towards the low density regime. This is necessary due to fluctuations especially in the early phases of the simulation.
The internal energy is constructed from the temperature through the assumption of an ideal gas, $\rm 3/2\ k_{B} T$, and is reduced by the properly weighted energy released by molecule formation and recombination of ions. Following the procedure of \texttt{MARCS} \citep{Gustafsson1975,Gustafsson2008}, only molecules containing H+, H2 and H2+ are considered for the summation of $\rm e_{int}$.
The resulting table is then interpolated to a $\rm e_{int}-\rho$ grid, which is chosen to be equally spaced in the natural logarithm as well. This final change of grids is necessary due to the appearance of the energy rather than temperature in Eq. \ref{eq:energy}, which means that as the simulation advances energy and density in time, temperature is a dependent variable and needs to be provided by the EoS rather than being an input parameter. We note however, that due to non-linearities introduced by molecule formation and destruction, as well as ionisation edges, the sampling of this final $\rm e_{int}-\rho$ table needs to be significantly higher to resolve step transitions in the $\rm e_{int}-T$ relation. We found that specifically towards the surface layers of the sun ($\rm T \lesssim 7\,000$K), the difference between opacities obtained from tables on $\rm  e_{int}$ and $\rm  T$ grids reach values beyond $\rm 20 \%$. We avoid this issue by up-sampling the table to a resolution of $\rm 1024$ points in $\rm e_{int}$, which results in a deviation of $\rm \lesssim 1 \%$. Because interpolation in the EoS table does not contribute significantly to the overall computation time, this up-sampling has negligible impact on performance.
\subsection{Opacity binning} \label{subsec:opacities}
\subsubsection{Procedure}
Because monochromatic radiative transfer is a computationally expensive task we follow the well known strategy of opacity binning. Following the procedure presented in \cite{Nordlund1982}, which has been adopted as a basis for different RHD calculations \citep[e.g.][]{Magic2013a,Vogler2004}, we sort a set of monochromatic opacities in bins, the selection of which is further described in the following section.

In each bin $\rm i$ we compute the mean opacity $\rm\kappa_i$ using a weighted transition from the Rosseland mean $\rm\kappa_{ross, i}$ in the interior to the Planck mean $\rm\kappa_{Planck, i}$ in the optically thin layers \citep{Nordlund1982}.
We then interpolate between the two limits using the optical thickness $\rm w_{thick}$ of the corresponding point in the table
\begin{align}
	w_{thick} &= 1 - e^{-2\,\tau_{ross, i}}, \label{eq:wthin}\\
    \kappa_i &= \kappa_{Planck, i}\ (1 - w_{thick}) + \kappa_{ross, i}\ w_{thick}.
\end{align}

This formulation is similar to the expression used in \citet{Magic2013a}, however differs by the use of the Planck mean opacity instead of the intensity mean. The expression is also equivalent to the one developed in \citet[][their eq. 3.51]{Ludwig1992}, defined as 
\begin{equation}
    \kappa_i = 2^{\frac{-\tau_i}{\tau_{1/2}}}{\kappa}_{ planck, i} + \left( 1-2^{\frac{-\tau_i}{\tau_{1/2}}} \right) {\kappa}_{ross, i},
\end{equation}
with $\rm \tau_{1/2}$ defined as the turnover depth where Planck and Rosseland mean opacity are weighted equally.
Whether one or the other is more relevant depends on if absorption (of intensity) or emission (of Planck source) is the more important effect. But in practice it may make little difference, and the Planck mean is more convenient to compute. To compute the Rosseland optical depth $\rm \tau_{ross, i}$ in Eq. \ref{eq:wthin} for each point in the table we follow the approximation from \cite{Ludwig1992}, that is
\begin{equation}
    \tau_{ross, i} = \kappa_{ross, i} \cdot \frac{p}{g}\ ,
\end{equation}
which originates from the equation of hydrostatic balance.

The radiative heating as given in Eq. \ref{eq:heating} furthermore requires knowledge of the source function $\rm S_{\nu}$, which in LTE is given by the Planck function $\rm B_{\nu}$ and hence only depends on the local temperature and frequency $\rm \nu=c/\lambda$, where $\rm c$ is the speed of light. The integral in Eq. \ref{eq:heating} can be split up in contributions from each bin as
\begin{align}
q_{rad} &= 4\pi \rho \sum_i \int_{\lambda(i)} \kappa_{\lambda} \left( J_{\lambda} - S_{\lambda}\right) \ d\lambda\ . \label{eq:heating_binned_1}
\end{align}
In the approximation of opacity binning, the mean opacity of each bin is assumed for the whole set of wavelength points belonging to this bin, and hence can be extracted from the integrand in Eq. \ref{eq:heating_binned_1} such that
\begin{align}
q_{rad} &= 4\pi \rho \sum_i  \kappa_{i}\int_{\lambda(i)} \left( J_{\lambda} - S_{\lambda}\right) \ d\lambda \\
	&=  4\pi \rho \sum_i  \kappa_{i} \left(  \int_{\lambda(i)}  J_{\lambda}\ d\lambda  - \int_{\lambda(i)} S_{\lambda}\ d\lambda \right) \ . \label{eq:heating_binned_2}
\end{align}
The computation of the mean intensity requires the solution of equations of the following form, depending on the angle $\rm \mu = cos(\theta)$, 
\begin{align}
I_{\nu, \mu} = \int S_{\nu} e^{-\tau_{\nu}/\mu} \ d\tau_{\nu}/\mu \label{eq:radiative_transfer},
\end{align}
where  $\rm J_\nu = 1/(4\pi) \int_{4\pi} I_{\nu} d\omega$ is the intensity $\rm I_\nu$ averaged over solid angle $\omega$ and $\rm d\tau_{\nu} = \rho \kappa_{\nu} dz$ is the optical depth increment at frequency $\rm \nu$ and geometrical depth increment $\rm dz$. In this work we rely on four $\rm \phi$ (azimuthal) and two $\rm \mu$ (polar) angles, and two additional angles with $\rm \mu=\pm1$, so in total ten different angles chosen in a Gauss-Radau quadrature. The selection of angles is equivalent to the \texttt{STAGGER} approach. In \texttt{CO$^5$BOLD} usually a higher number of angles is used \citep[see e.g.][]{Beeck2012}, but it depends on the particular application of the code. In the binning approximation the optical depth is computed from the binned opacity, and hence independent of  $\rm \lambda$, which means the optical depth terms in Eq. \ref{eq:radiative_transfer} can be pulled out of the $\rm \lambda$ integral in Eq. \ref{eq:heating_binned_2} such that
\begin{align}
	\int_{\lambda(i)}  J_{\lambda}\ d\lambda &\sim  \int_{\lambda(i)} \left( \iint S_{\lambda} e^{-\tau_{i}/ \mu}\ d\tau_{i}/\mu \ d\omega\right) \ d\lambda \\
    &\sim   \iint  e^{-\tau_{i}/ \mu}\ d\tau_{i}/\mu \ d\omega \int_{\lambda(i)} S_{\lambda} \ d\lambda . \label{eq:binned_intensity}
\end{align}

The integration over wavelength points corresponding to bin $\rm i$ can hence be executed before the solution of the radiative transfer. The resulting binned source function 
\begin{equation}
S_i = \int_{\lambda(i)} S_{\lambda} \ d\lambda, \label{eq:source_function_binned}
\end{equation}
can thus be pre-computed, included in the opacity table, and used together with the binned opacity to obtain the radiative heating during the simulation, with reduces the computational burden significantly. 

Note that the monochromatic source function in NLTE would depend on the radiation field, which strictly speaking would require knowledge of the model atmosphere prior to its simulation. However, NLTE calculations for multiple species are extremely computationally expensive, especially in 3D, which is why all 3D RHD codes, including \texttt{STAGGER}, \texttt{MURaM}, and \texttt{CO$^{5}$BOLD} generally adopt the LTE approximation for structure calculations. The effect of NLTE on the physical structure of models was investigated in other studies \citep[e.g][]{Short2005, Haberreiter2008, Young2014}, based on 1D HE modelling. These studies showed that the effect is very small, in particular for the Sun not exceeding 50 to 100 K in the atmospheric layers $-3 \lesssim \opd \lesssim +0.5$ \citep{Short2005}, whereas outside these layers the presence of the chromosphere or convection is much more important.
Including NLTE in structure computations would introduce additional dimensions to the opacity table and increase the complexity, which is why we for simplicity work in the LTE approximation. The spectrum synthesis on the relaxed 3D model atmosphere (see Sect. \ref{subsec:spectrum_synthesis}) however is can be done in NLTE.
\subsubsection{Selection of opacity bins}
\begin{figure}
	\includegraphics[width=0.5\textwidth]{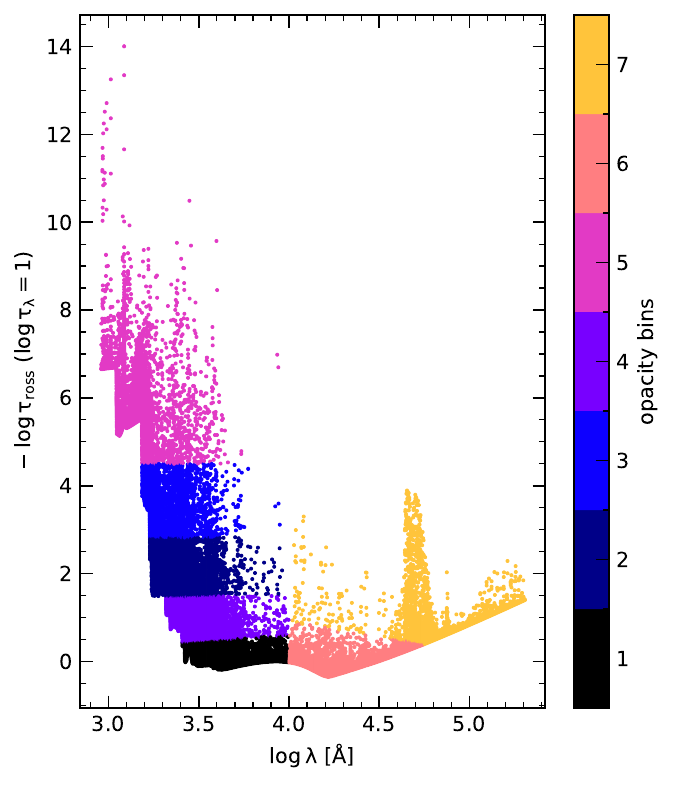}
	\caption{Formation height as a function of wavelength for monochromatic \texttt{MARCS} opacities, computed for an average 3D \texttt{STAGGER} solar snapshot.}
\label{fig:binning_marcs}
\end{figure} 

While there is no consensus among related studies on the exact number and positioning of opacity bins, there is a general understanding of the need to distribute bins to try to cover the regions most important for the energy balance. The \texttt{STAGGER} grid \citep{Magic2013a} is computed using 12 opacity bins, which were defined depending on their wavelength and formation height on the Rosseland optical depth scale ($\rm \tau_{ross}$), that is the Rosseland optical depth where their monochromatic optical depth equals unity. The computation of these optical depth scales requires some initial knowledge of the average atmospheric structure of the 3D model that is intended to be simulated. For the \texttt{STAGGER} grid an iterative procedure is involved at this stage, which adjusts the bin dimensions repeatedly as the simulation evolves in time. As a metric for the goodness of the binning they furthermore compute the monochromatic heating rates and compare them to the those of the binned result. They report typical differences of 
$\rm \max \left|q_{rad}^{bin}-q_{rad}^{\lambda}\right| / \max \left|q_{rad}^{\lambda} \right| \sim 3\%$ for their solar model, however they are able to improve this difference by further fine-tuning \citep{Magic2013a}. It remains to be tested how this translates to uncertainties of the 3D structure of the final model, and especially on the observables, for example the stellar spectrum. In the case of our solar model, our binning method as described below archives $\rm 2.8 \%$ in the same metric.

A very similar binning is applied in the \texttt{CO$^5$BOLD} \citep[e.g.][]{Freytag2012,Ludwig2023} simulations, where however the bin selection involves less fine-tuning as they are placed in regular optical depth and wavelength intervals \citep{Beeck2012}. The \texttt{MURaM} solar model \citep{Vogler2005} goes one step further and relies on only four opacity bins, which are wavelength independent and chosen based only on the monochromatic optical depth \citep[as in][]{Nordlund1982}. There is no further optimisation procedure involved. 

Despite their drastic differences in bin selection, it has been shown in \cite{Beeck2012} that the relaxed simulations of the solar photosphere are remarkably similar in terms of the mean temperature and density stratification. The thermal structure of the solar models presented in \citeauthor{Beeck2012} furthermore show a good resemblance on the optical depth scale (see their Fig. 10, 11) with mean differences below $\rm 100 \ K$ in the regions relevant for spectral line formation, which suggests a rather small influence of the opacity binning procedure on the stellar spectrum.

In this work we apply a simple, automatic binning procedure and do not alter the binning in the course of the simulation. We select seven opacity bins, which are distributed similarly to the bin distribution employed by \cite{Magic2013a}. The number of bins has been chosen as a compromise between accuracy and computational efficiency. In our simulations between $60$ and $80 \%$ of the entire computation time is spent solving the RT, a reduction of opacity bins hence has a very significant effect.
We first separate the strong lines at shorter wavelength from the redder part of the spectrum. In the blue part, we furthermore specify an upper limit on formation height, beyond which we collect all wavelength points in a single, high opacity bin in order to assure sufficient sampling in the regime closer to the continuum.
The formation height edges of the remaining blue bins are then selected automatically using a generic k-means clustering approach \citep[e.g.][]{Hartigan1979}. 
The red wavelength points are binned accordingly. Following this procedure, we split the most relevant opacity sources below the Rosseland formation depth of $\rm \rm -4.5$ in the blue part of the opacity table in four bins, which leaves two bins for the red part of the spectrum and one bin collecting the remaining strong lines.
This procedure is deliberately designed to be as generic as possible, while at the same time producing consistent results for variety of different stars without human intervention. It can be executed automatically prior to the simulation by using an initial 1D model (for example a 1D HE model) and does not involve an iterative procedure. Yet, it captures the location of opacity nodes in the $\rm log \lambda - log \tau_{ross}$ diagram through clustering and hence provides more bins where a higher resolution in required. We have tested the approach by exploring the differences to the monochromatic heating rate with the heating computed using four, five, seven, eight, and 12 bins with parameters similar to those chosen by other authors. We find that for all these combinations the differences relative to the cooling peak are at a sub-percent level. Therefore we consider our chosen scheme as appropriate within the context of our work, as a  compromise between very expensive fine-tuning \cite[\texttt{STAGGER,}][]{Magic2013a} and wavelength independent \cite[\texttt{MURaM,}][]{Vogler2004,Vogler2005} approaches. 

For the solar model presented in this work, we bin tabulated opacities generated by the the \texttt{MARCS} code using the chemical composition presented in \cite{Magg2022}. The formation height of each wavelength point in this table is computed using the average solar \texttt{STAGGER} model. The corresponding binning is presented in Fig. \ref{fig:binning_marcs}, where we show the formation height as a function of wavelength and indicate the bin assignment with colours.
\subsection{Initial \& boundary conditions} \label{subsec:initial_conditions}
Every model simulated in this works relies on only three numbers that will determine its final state; the temperature and density at a single point within the atmosphere, as well as the surface gravity, $\rm \log(g)$. This initial seed can be placed at any height $\rm z_0$, which should be chosen such that the zero-point of the resulting z scale roughly coincides with the optical surface. From here, an initial 1D HE model is constructed by integrating adiabatically from the initial seed upwards and downwards with a constant step in $\rm \ln \rho$.
Considering a fluid parcel of constant mass, the change of its internal energy per unit mass $\rm e_{int}$ is given by the first law of thermodynamics
\def\d{{\rm d}}
\begin{equation}
\d e_{int} = T \d \emph{S} - p\,{\rm d}V = T \d \emph{S} - {p}{V}\,\d\ln V = T{\rm d}\emph{S}  + \frac{p}{\rho} \, \d\ln{\rho} ,
\end{equation}
where $\rm \emph{S} $ is the entropy per unit mass and $\rm V$ the volume of the fluid parcel. If the system is adiabatic, with $\rm \d \emph{S} =0$, then the above equation links a change in log density directly to a change in internal energy per unit mass, via a prefactor $\rm p/\rho$ that behaves similarly to the temperature $\rm T$.
For a given step in $\rm \ln{\rho}$ the internal energy per unit mass can thus be computed iteratively, with a procedure that converges rapidly.  The z-scale is constructed from hydrostatic balance
\begin{equation}
  	{\rm d}z = -\frac{p}{\rho g}\,\d \ln p.
\end{equation}

Towards the surface the adiabatic structure is terminated once a certain minimal energy is reached, which is kept fixed for the remaining integration. 
The resulting 1D structure is then interpolated onto the coordinate grid of the 3D simulation domain, assuming no variations in horizontal directions. 

The initial velocity field is initialised as a Gaussian drop-off with distance from $\rm z=0$,
including a random factor to kick-start convective motions.
The entropy at the bottom of the atmosphere is kept fixed during the remaining parts of the simulation and serves as a bottom boundary condition.

Alternatively, it is also possible to skip the adiabatic initialisation and start directly from any 1D model atmosphere (either a 1D HE or a horizontally averaged 3D model), where the internal energy is computed consistently with the EoS that will be used in the simulation. This procedure is especially useful when a specific bottom boundary condition is required, for example if a comparison with other simulations is desired. All models generated in this work are initialised using an average 3D (i.e. 1D) \texttt{STAGGER} model atmosphere to make the comparison more straightforward. We note however, that also in this case only three numbers, density and temperature at the bottom of the initial model and surface gravity, determine the ultimately emerging 3D structure.

The fixed entropy is enforced by requiring inflows to be isentropic, while keeping outflows unchanged. At the top of the atmosphere, the energy per unit mass is kept constant, while density is assumed to drop off exponentially with pressure scale height. The horizontal velocity field is furthermore decaying with the same factor, equivalent to a non-shear boundary, while the vertical velocity is dropped off on half a scale height. The radiation boundary, that is the surface heating term, is estimated from the optical depth in the vertical, and zero in the horizontal. The surface gravity is furthermore kept fixed and is the same throughout the simulation domain.

\subsection{Relaxation} \label{subsec:relaxation}
The relaxation procedure consists of three stages, which are mainly characterised by the cooling sources and the presence of damping terms. The first phase starts from the initial model described in the previous section. In order to start the convective motions quickly and reliably the only source of cooling at this early stage of the simulation is a strong, artificial, so-called 'Newton cooling'. It effectively removes (or adds) heat to the system so that its internal energy reaches a given value $\rm e_0$ on a given timescale $\rm \delta t_N$ at a given position $\rm s_0$ with a scale of $\rm s_{1}$. The resulting heating per unit volume $\rm q_N$ is given as
\begin{align}
		q_{N} &= (e_{0}-e)  \frac{f}{(1+f) \cdot \delta t_N}, \\
        f &= e^{- \frac{s-s_0}{s_1}}.
\end{align}
There is no radiative cooling at this stage. The average vertical component of the velocity field $v_z$ is furthermore damped on the timescale $\rm \delta t_f$ via the addition of a friction term of magnitude $\rm \rho \, v_z / \delta t_f$ to damp out artificial radial p-mode oscillations -- excited by the rapidly changing thermal structure -- that would otherwise survive for a long time.  Because of approximate conservation of wave mode energy, even weak wave modes initiated near the bottom boundary can grow in amplitude when travelling towards the surface, and can ultimately even become disruptive. This initial procedure has emerged from countless tests and provides a reliable and fast pathway from the initial, adiabatic cube to the first semi-realistic 3D model atmosphere that can be evolved further in time in the following phases \cite[e.g.][]{Freytag2012}. This intermediate model already has developed convective motions and displays a granular pattern towards the surface layers. 

We note that the initial Newton cooling phase is strictly speaking not required, as the radiative transfer itself is capable of providing the relevant cooling to develop the radiative surface. However, the advantage of an initial Newton cooling phase is that (i) its strength is a hyper-parameter that can be optimised to achieve a more realistic velocity field in a shorter time, and (ii) it does not require the solution of the radiative transfer equation, which is expensive in terms of computation time and potentially unstable on an initially chaotic atmospheric background. The particular choice of the additional hyper-parameters is irrelevant for the final structure of the model, but numerical stability is enhanced if the cooling surface at $\rm e_0$ lies close to the true optical surface that will emerge from the actual radiation field. However, the correct surface will emerge in the second phase regardless of the exact decision. 

We chose the hyper-parameters $\rm e_0$, $\rm s_0$, $\rm s_1$ and $\rm \delta t_{N}$ based on the 1D HE model that is used for initialising the 3D cube. After using the opacity table to compute the Rosseland optical depth scale of the initial 1D HE model we place the upper boundary at $\rm  \log \tau_{ross}\sim -5.5$. From here we pick the physical height, density, and temperature and use the EoS to compute the internal energy at this point in the atmosphere. This directly provides estimates for $\rm e_0$ and $\rm s_0$. A qualitative investigation has shown that a scale of $\rm  s_1=0.1\ Mm$ and timescale of $\rm \delta t_{N}=10\ s$ produces adequate results in the solar case. For stars other than the sun we scale those parameters according to changes in physical dimensions of the star and strength of their velocity field using the courant condition.

\begin{figure*}[ht]
\includegraphics[width=\textwidth]{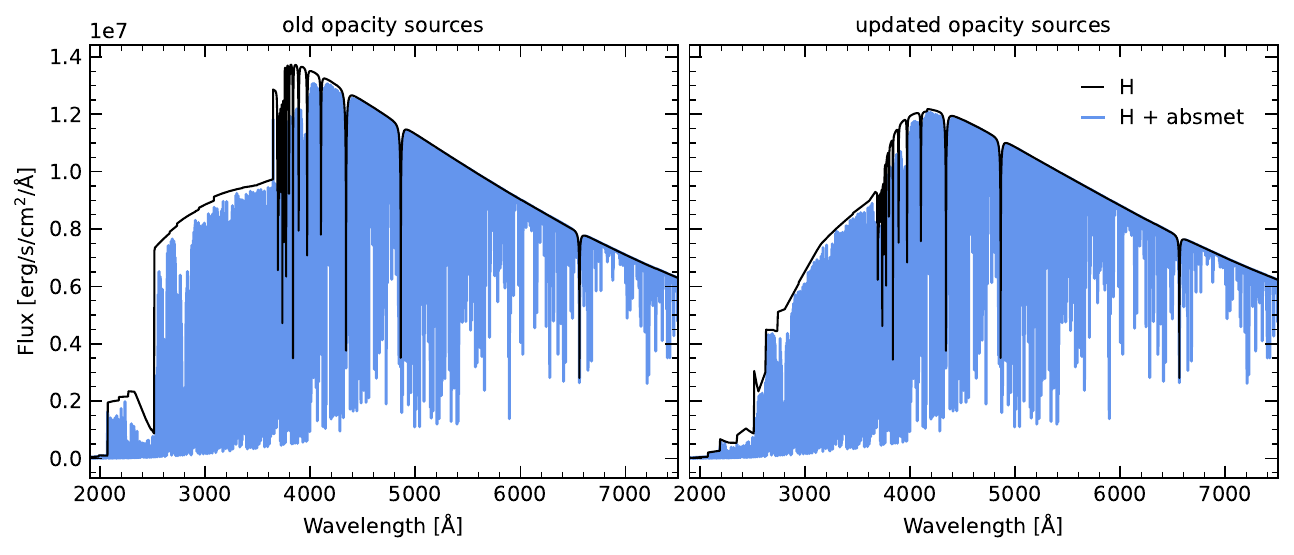}
\caption{Model spectral energy distribution (SED) of the Sun computed using \texttt{MULTI3D@DISPATCH} and the solar \texttt{M3DIS} snapshot. Left and right panels show the SEDs computed with the old and our new updated background bound-free opacities. The occupation probability formalism \citep{Dappen1987} is used in the updated opacities for H I bound-free transitions. The black line shows the result of running H lines in LTE on top of the LTE background bound-free opacities. The blue line shows the model flux when tabulated LTE opacities of bound-bound radiative transitions are included.}
\label{fig:SunSED}
\end{figure*} 
\begin{table}
\centering                          
\caption[]{Updated bound-free and free-free opacities and their respective sources. A sub-set of the \texttt{MARCS} and \texttt{Turbospectrum} opacities \citep{Gustafsson2008, Plez2012, Gerber2023}.}
\label{tab:ContOpac}
\begin{tabular}{l l l}            
\hline\hline                        
Species   &  Type &  Source \\           
\hline                              
H                                               &  bf, ff   & \cite{Karzas1961} \\
H$^{-}$                                         &  bf       & \cite{McLaughlin2017} \\
He, C, N, O, Mg                                 &  bf       & TOPbase \citep{Seaton1994} \\
 Al, Si, Ca, Fe                                 &  bf       & TOPbase \citep{Seaton1994} \\
CH, OH                                          &  bf       & \cite{Kurucz1987} \\
H$^{-}$                                         &  ff       & \cite{Bell1987} \\
He, C, Mg, Si                                   &  ff       & \cite{Peach1970} \\
He$^{-}$                                        &  ff       & \cite{John1994} \\
H$_{2}^{-}$, O$^{-}$, CO$^{-}$, H$_{2}$O$^{-}$  &  ff       & \cite{John1975a, John1975b} \\
H \ur{1} + H \ur{1}                             &  ff       & \cite{Doyle1968} \\
H$_{2}^{+}$                                     &  ff       & \cite{Mihalas1965} \\
\hline
\end{tabular}
\end{table}

This early atmosphere is advanced in time until a granular pattern has emerged. For the solar model, the duration of the first phase corresponds to roughly $\rm 30-40$ stellar minutes. After this, in the second phase, Newton cooling is gradually reduced and replaced by radiative cooling. Newton cooling is reduced by decreasing the cooling rate exponentially, with a decay time scale of two stellar minutes, until it reaches 1\% of its original value. After this, it is turned off entirely.

The switching of the cooling source is accompanied by additional, destabilising oscillations, which is why the friction term remains present during this phase. We hence wait an additional $\rm \sim 20$ stellar minutes (until $\rm 50$ stellar minutes total simulation time) before we start to fade out the friction term in the same manner as described above, but with a longer characteristic time scale of $\rm 15$ minutes.

\dis uses local timestepping, where the local timestep is derived according to the local CFL condition \citep{Courant1967}. During the radiative transfer, this usually results in a timestep between $\rm 0.1s$ and $\rm 0.4s$. The radiative heating rate is included in the courant condition directly through the velocity $\rm u = ds \cdot q_{rad}/p$, with grid spacing $\rm ds$.
\subsection{Averaging of 3D RHD cubes} \label{subsec:averaging}
For better visualisation and comparison purposes we average the full 3D data cubes to averaged 1D vertical profiles. Since the absolute height scale in Cartesian coordinates bares no physical meaning, especially in the context of spectrum synthesis, the averaging of the data cubes is made on the Rosseland optical depth scale $\rm \tau_{ross}$. For this, we proceed as follows.

First, the Rosseland opacity for every point in the 3D cube is looked up in the EoS table. From this, $\rm \tau_{\rm ross}$ is computed via column-wise integration. The entire cube is then interpolated to an optical depth scale equidistant in natural logarithm, again column-wise. The result is a 3D cube, where each horizontal plane corresponds to a common optical depth. The averaging of different quantities is then done per plane.

In order to be entirely consistent and avoid controversy originating from different opacities and integration limits of the Rosseland opacity, we apply this same procedure also to the reference models. This means that all models present in this paper are on the same Rosseland height scale and consequently averaged in the same way. The average models are used for the sole purpose of comparisons and for gaining an intuitive understanding of the physical structure of models. We do not rely on averaged models for the calculations of synthetic observables.
\subsection{Spectrum synthesis} 
\label{subsec:spectrum_synthesis}
Calculations of high-resolution synthetic spectra presented in this work are carried out using the \texttt{MULTI3D} code \citep{Leenaarts2009}, which we have also ported into the \dis framework. For details about the code we refer to  \citeauthor{Leenaarts2009} and minor updates described in \citet{Gallagher2020} and \citet{Bergemann2019}. 
In summary, the code employs the method of short characteristics and accelerated lambda iteration \citep{Rybicki1992} to solve the equations of radiation transfer and statistical equilibrium. After completing the final NLTE iteration, the outgoing flux is calculated with the method of long characteristics and the Lobatto quadrature. To enable detailed 3D spectrum synthesis, we have updated the code to include an extensive collection of tabulated bound-free and free-free cross-sections and partition functions for 92 elements. The partition functions are taken from \cite{Irwin1981}, while the sources of the continuous opacities are listed in Tab. \ref{tab:ContOpac}. The occupation formalism is adopted for H I only, following the \texttt{HBOP} routine by \cite{BarklemPiskunov2003}. This update ensures that the opacity sources of \texttt{MULTI3D} match those used in  \texttt{Turbospectrum} \citep{Plez2012, Gerber2023}. As a result, \texttt{MULTI3D} can now incorporate bound-bound radiative opacities from comprehensive linelists provided in the Turbospectrum format. Alternatively, users can include pre-computed opacity tables to account for line blanketing. This allows for the use of fully self-consistent line opacities in both the 3D RHD modelling and the spectrum synthesis. A detailed description of all updates will be discussed in an upcoming paper (Hoppe, Bergemann in prep.) We furthermore include the detailed linelist assembled by the Gaia-ESO collaboration, as it provides carefully-assessed atomic data for the diagnostic lines of the majority of chemical elements, including the transition probabilities and damping \citep{Heiter2021}. For H line profiles we follow the \texttt{HLINOP} routine by \cite{BarklemPiskunov2003} which includes Stark broadening, natural broadening and fine-structure splitting. The self-broadening is treated using the BPO theory in the case of the Balmer lines \citep{BPO2000} and using the theory by \cite{AliGriem1965} in all other cases. In order to ensure backward compatibility with Turbospectrum we have implemented the same treatment of elastic H collisions (with four different options) as in the latter \citep[see][and documentation to \texttt{Turbospectrum2020}\footnote{Publicly available at \url{https://github.com/bertrandplez/Turbospectrum_NLTE}}]{Gerber2023}.

Fig. \ref{fig:SunSED} shows the impact of the new (right panel) versus old (left panel) tabulated bound-free opacities, as well as the use of the occupation probability formalism. The latter explains the smooth transition in the flux continuum around 4000 \AA, where there was a clear cut when classical Debye shielding was used instead. The updated continuous opacities are responsible for the depression in the UV where the flux has decreased noticeably. The slope of the continuum towards the infra-red has mostly been affected with the updated H$^{-}$ opacities from \cite{McLaughlin2017}.

\begin{figure}
\includegraphics[width=\columnwidth]{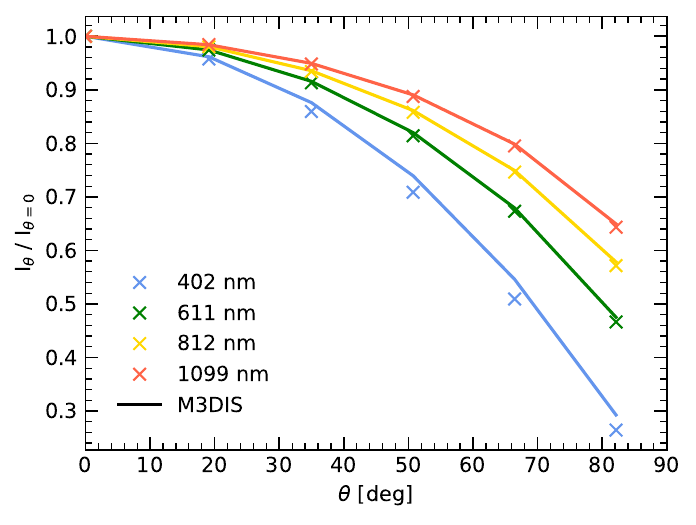}
\caption{Angle dependent (centre-to-limb) intensity variation of solar \texttt{M3DIS} snapshot at 4 discrete wavelengths. The solar spatially-resolved intensity data from \cite{Neckel1994} are shown with crosses. We note that in the latter paper, only polynomial fits are available, here for didactic reasons we present them as discretised points.}
\label{fig:CLV}
\end{figure} 

Fig. \ref{fig:CLV} shows the centre-to-limb variation (CLV) of one solar \texttt{M3DIS} snapshot compared with the observations by \cite{Neckel1994}. There is excellent agreement between the synthetic data and the observations at 611, 812 and 1099\,nm, while slight differences are seen in the blue 402\,nm. The same offset is seen in an alternative 3D RHD solar model from \texttt{STAGGER} which was also used in \citet{Bergemann2012}, hence it unlikely due to the intrinsic properties of the 3D RHD models. Offsets of a very similar amplitude are also seen at 400\,nm in other 3D RHD models, such as MURAM \citep[][for a detailed comparison]{Beeck2012}. Comparing with \citet[][their Fig. 3]{Pereira2013}, we find that their 3D RHD models show very small yet discernible discrepancies around 402 nm, especially at larger angles $\mu=0.5$ and $0.3$, and similarly to our results the intensity ratios predicted by the 3D model are systematically higher compared to those of the observed data. This may either indicate limitations of the observed data in the blue or the effects of a magnetic field and a chromosphere that are not included in the standard 3D RHD simulations \citep{Carlsson2019}.
\section{Results} \label{sec:results}
\subsection{Influence of spatial resolution on atmospheric structure}\label{sec:resolution}
To allow the atmospheric models to converge towards a realistic thermodynamic structure the spatial resolution of the simulation domain, and especially the vertical resolution around the optical surface, are of utmost importance. In particular, the resolution adopted for the radiative solver is crucial to properly sample the cooling peak around the optical surface. We note that in this section, the discussion will refer to the spatial resolution only.

In order to investigate this effect we compile a comparison between three models originating from identical initial conditions, but with varying spatial resolution. The adopted HD grid spacing is $\rm 25.6\,km$ for the low and intermediate resolution model, and $\rm 15.4\,km$ for the high-resolution model. The vertical grid for the RT solver is denser by a factor 2 for the intermediate and high-resolution models, with a spacing of 12.8 and 7.7\,km respectively. The low resolution model has the same number of points in HD and RT. We summarise the different resolutions in Tab. \ref{tab:resolution}.

Due to the modular nature of \texttt{DISPATCH} it is straight forward to solve the radiative transfer on a grid different from the grid on which the underlying hydrodynamic solver operates. Tasks that are commissioned with solving Eq. \ref{eq:heating} receive the necessary information about the current structure of the atmosphere, $\rm \rho$ and $\rm e_{int}$, via linear interpolation from tasks that are responsible for solving Eq. \ref{eq:continuity}-\ref{eq:energy}. The final radiative heating rate $\rm q_r$ is then interpolated back to the initial HD grid and added to the energy balance for the next update of variables.
\begin{table}
\centering                          
\setlength{\tabcolsep}{3.5pt}
\caption[]{Resolution overview of \texttt{M3DIS} solar models.}
\label{tab:resolution}
\begin{tabular}{l l c c c c}            
\hline\hline                        
\smallskip
        & & $\rm N_x \times N_y \times N_z$ (HD) & $\rm N_z$ (RT) & $\Delta z$ (RT) & $\rm N_{CPU}$  \\           
        & &   & &  km & hours \\           
\hline                              
 A & low           & $\rm 180 \times 180 \times\ 90$  & $ 90$ & 25.6 & $\rm 226$ \\
 B & interm.        & $\rm 180 \times 180 \times\ 90$  & $ 180$ & 12.8 & $\rm 576$ \\
 C & very high*  & $\rm 390 \times 390 \times 195$ & $ 390$ & 5.9 & $\rm 15\,206$\\
  & & & & \\
 D & high (ref.) & $\rm 300 \times 300 \times 150$ & $300$ & 7.7 & $\rm 5\,242$ \\\hline
\end{tabular}\\
\tablefoot{
    \tablefoottext{*}{Due to the high resolution of this model, the stellar time it evolved within 24h wall-clock time is reduced compared to models A, B, and D.}
    
    For all models the number of grid points in each dimension ($\rm N_x$, $\rm N_y$, $\rm N_z$), as well as the vertical grid spacing ($\rm \Delta z$) and required CPU time ($\rm N_{CPU}$) is shown. The physical size of the box in all simulations is $\rm 4.6 \times 4.6 \times 2.3\ Mm$. The (x, y) resolution of the radiative transfer (RT) solver is the same as that for the hydrodynamic (HD) solver, and is not shown here to avoid redundancy. See text.
}    
\end{table}
\begin{figure}
	\includegraphics[width=\columnwidth]{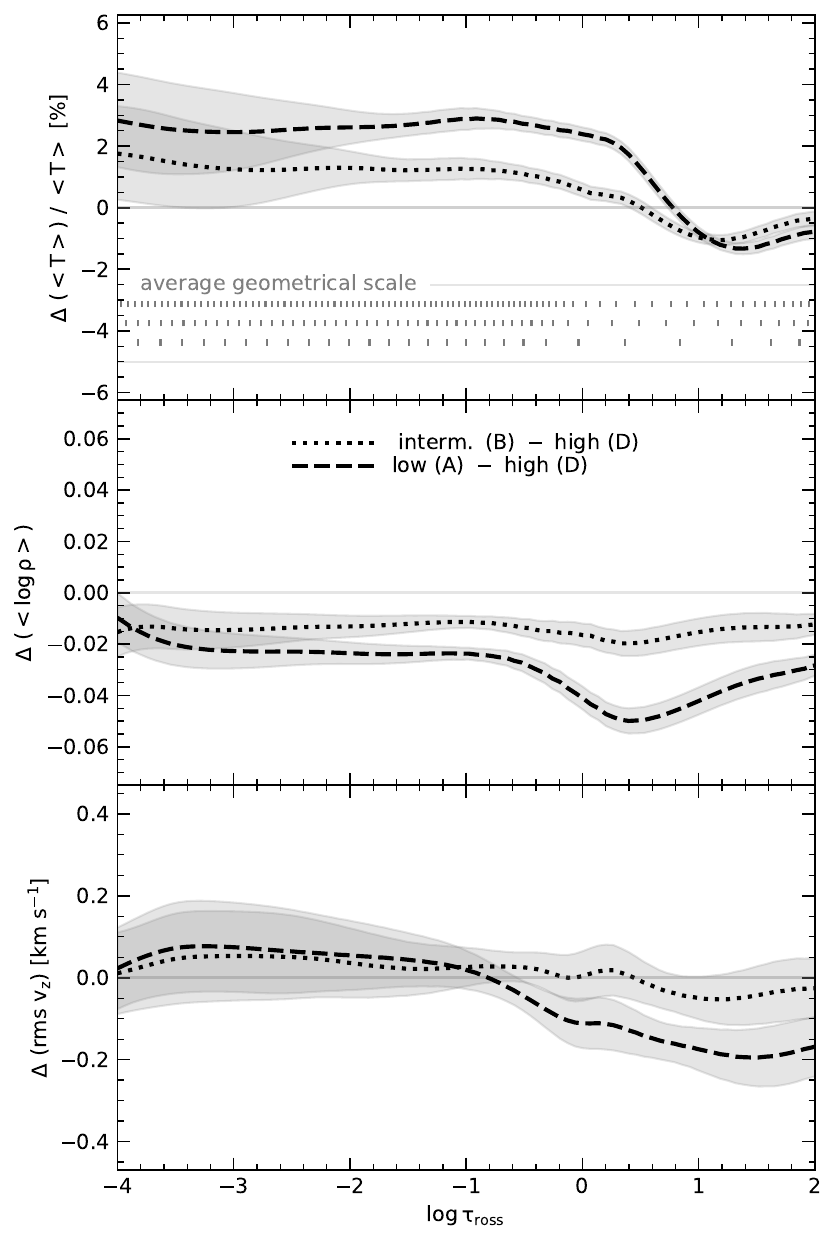}
	\caption{Average temperature (top) and density (middle) profiles, as well as rms vertical velocity (bottom) for models of different resolution. In each panel the difference to the highest resolution model ($\rm \Delta z = 7.7\ km$) is shown. Shading corresponds to the standard deviation between 9 different snapshots of the high resolution model. In the top panel we include the average geometrical depth scale of models A, B and D (from bottom to top) as vertical grey bars.}
\label{fig:hd_resolution}
\end{figure} 

The comparison between the resulting average profiles is shown in Fig. \ref{fig:hd_resolution}. The statistics is made on planes of constant Rosseland optical depth, as described in Sect. \ref{subsec:averaging} and then averaged over nine snapshots to obtain a robust comparison. The snapshots are taken every $200$ seconds in time. Our time sequence thus covers at least one convective turnover time, which, at the bottom of our simulation, amounts roughly to $\sim$ 5 to 7 minutes. We note that the convective turnover time can be estimated locally using 1D HE models (kindly provided to us by H.-G. Ludwig) yielding the estimates very similar to the ones above. Each panel shows the difference $\rm \Delta = av_1 - av_2$ between two average structures $\rm av_1$ and $\rm av_2$. Relative temperature differences in the upper panel are computed as $\rm \Delta = (av_1 - av_2) / av_2$. The shaded regions correspond to one standard deviation ($\rm std$) between different snapshots of the highest resolution model, that is $\rm av_1 - (av_2 \pm std_2)$ for the upper and lower limit. In case of relative differences $\rm (av_1 - (av_2 \pm std_2)) / (av_2 \pm std_2)$ are the limits of the shaded region.
By exploring the differences in the average temperature (top panel) and density profiles (middle panel), the effect of resolution becomes apparent mainly at the optical surface around $\rm \log \tau_{ross} \approx 0$ and towards the upper boundary of the star. In general, models of lower resolution are hotter and have a lower density in the upper layers. This behaviour is expected and directly related to the poor sampling of the cooling peak. In the relevant layers temperature and density change rapidly due to the sudden decrease in opacity. However, this transition region is spatially confined in the simulation domain and hence only covered by a small number of grid points. 
As a consequence the linear interpolation on this sparse grid overestimates the heating rate, which results in a shallower temperature decline and overall hotter upper atmosphere with lower densities. To illustrate the optical depth sampling around the optical surface we include the average optical depth derived from snapshots on the geometrical scale in the top panel of Fig.\ \ref{fig:hd_resolution}. Around $\rm \log \tau_{ross} \sim 0$ the step for low, intermediate, and high resolution models A, B, and D is $\rm 0.248$, $\rm 0.113$ and $\rm 0.062$ $\rm dex$, respectively. As a comparison, the optical depth step in \texttt{MARCS} models ranges between $\rm 0.1$ and $\rm 0.2$, depending on the location in the atmosphere (Plez, priv. conv.).

Between A ($\rm \Delta z=26\,km$) and D ($\rm \Delta z = 7.7\,km$) we estimate the magnitude of this effect to be on the order of $\rm 2.5\, \%$ on the average temperature stratification and between $\rm 0.02\,dex$ and $\rm 0.06\,dex$ for density. We furthermore note that the root-mean-square (rms) of the vertical velocity in the low resolution case on average is higher in the interior compared to the high resolution models. At the surface, the effect becomes minimal. We investigate in Sect.\,\ref{subsec:solar_spectrum} how these differences in the atmospheric structure translate to  observable quantities in the spectrum of the Sun.

For each resolution we furthermore include the computation time necessary to compute the model in the last column of Tab. \ref{tab:resolution}. Because solving radiative transfer is computationally expensive, the RT grid size has significant impact on the overall computation time. However, due to the unique speed provided by \texttt{DISPATCH} all models presented here are obtained within 24h\footnote{Computations were performed on the HPC system Raven and Cobra at the Max Planck Computing and Data Facility} from start to end, using only 1D HE models as initial conditions, as outlined in Sect. \ref{subsec:initial_conditions}.
We emphasise that the comparisons performed here are done with fixed bottom boundary conditions, and as a consequence the temperature differences illustrated in Fig.\ \ref{fig:hd_resolution} would result also in changes of the actual (as opposed to `label') effective temperature. In a grid-of-models context one would interpolate in the grid to the exact label effective temperature before computing synthetic spectra, which would most likely reduce the influence of resolution even further, since the changes shown in Fig.\ \ref{fig:hd_resolution} do not vary much in the optically thin layers.  It is thus likely that in a grid context the cost of modelling per grid point could be further reduced, relative to the results reported here. A summary of model properties and input physics is given in Tab. \ref{tab:stellar_parameters}. Here we compute the effective temperature by using the binned opacity table presented in Sect. \ref{subsec:opacities}.
\begin{figure}
\includegraphics[width=\columnwidth]{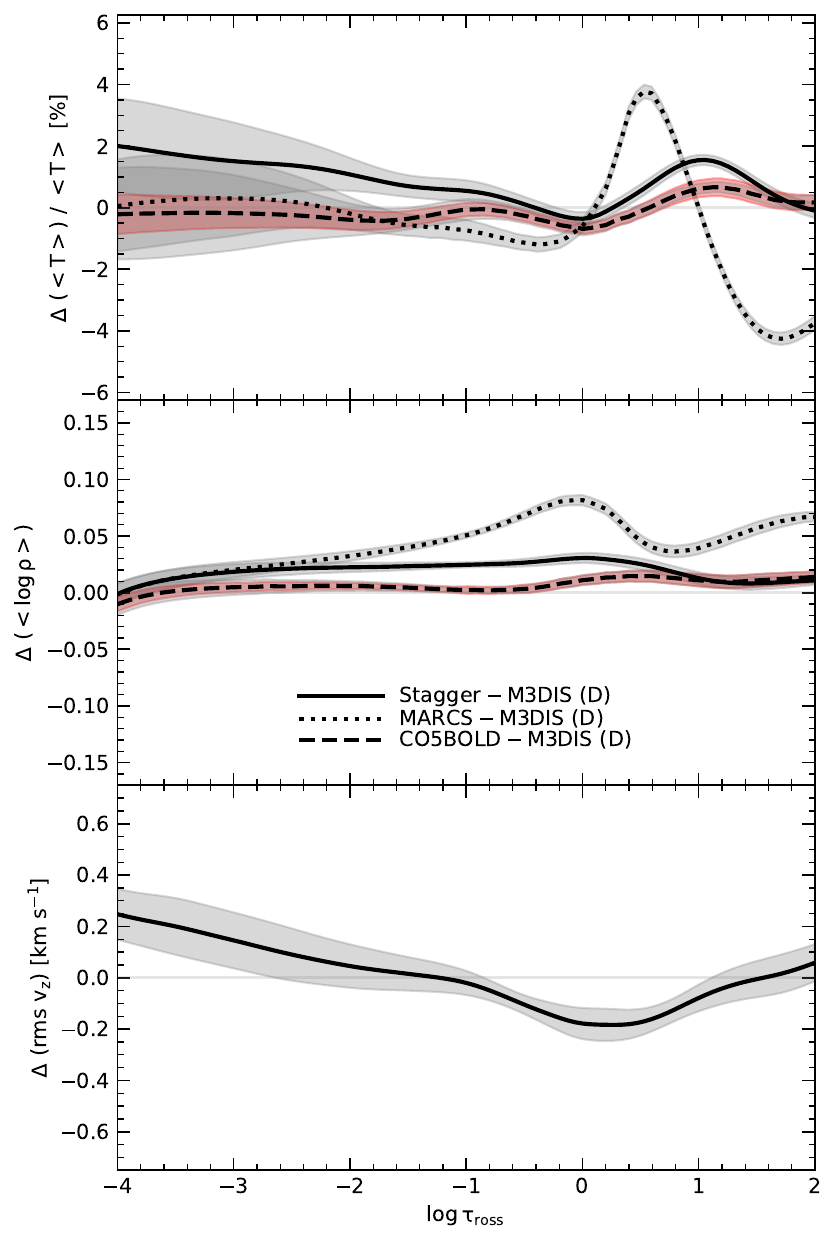}
\caption{Differences in average 3D temperature (top), density (middle) and rms vertical velocity (bottom) structure as a function of Rosseland optical depth for our best solar model (\texttt{M3DIS}), \texttt{STAGGER}, \texttt{CO$^5$BOLD} and \texttt{MARCS}. We include the \texttt{MARCS} models, because they are among the most widely-used 1D hydrostatic model atmospheres for FGKM type stars analyses. The average \texttt{M3DIS} structure and the grey shading correspond to mean and standard deviation between 9 different snapshots of the high resolution model, respectively. For \texttt{CO$^5$BOLD}, 
the mean of 20 different snapshots was used. We also include the standard deviation between these snapshots in red as a reference.}
\label{fig:model_comparison}
\end{figure} 
\begin{figure*}[ht]
\centering
\hbox{
\includegraphics[width=0.33\linewidth]{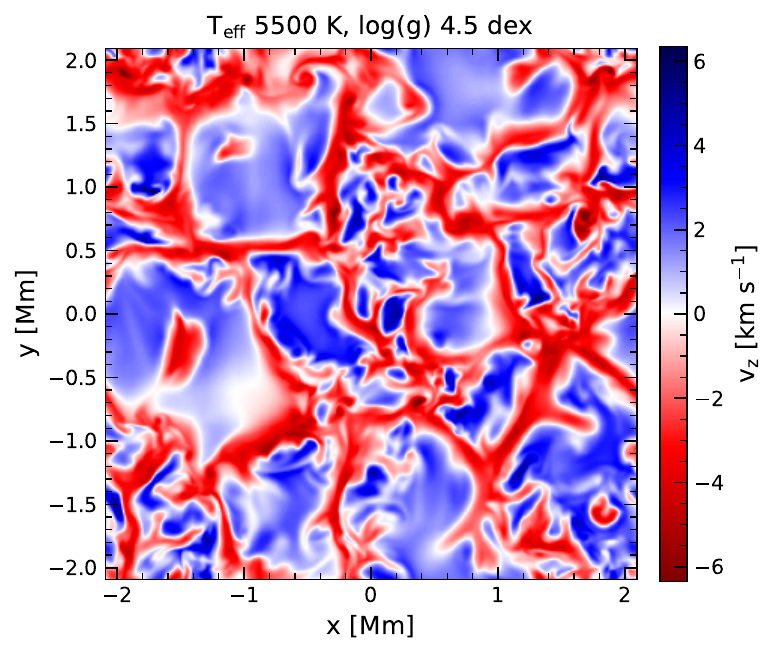}
\includegraphics[width=0.33\linewidth]{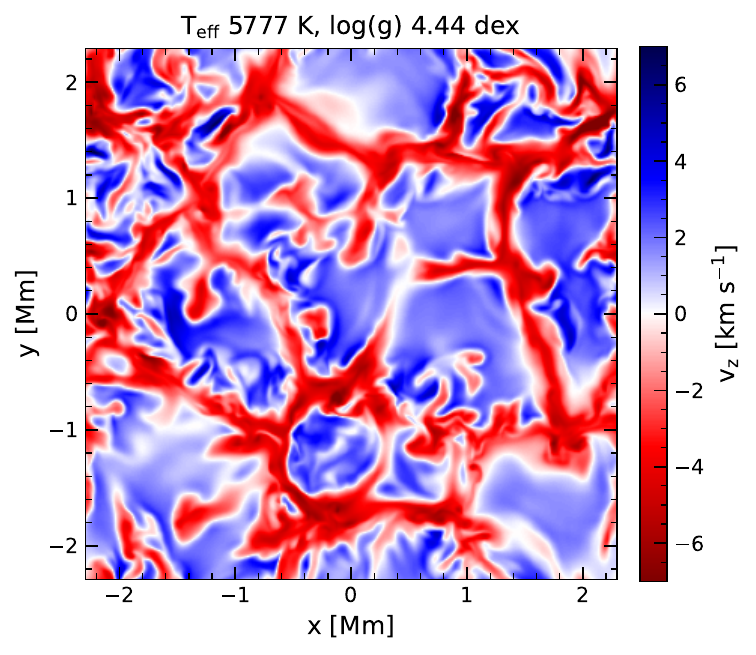} 
\includegraphics[width=0.33\linewidth]{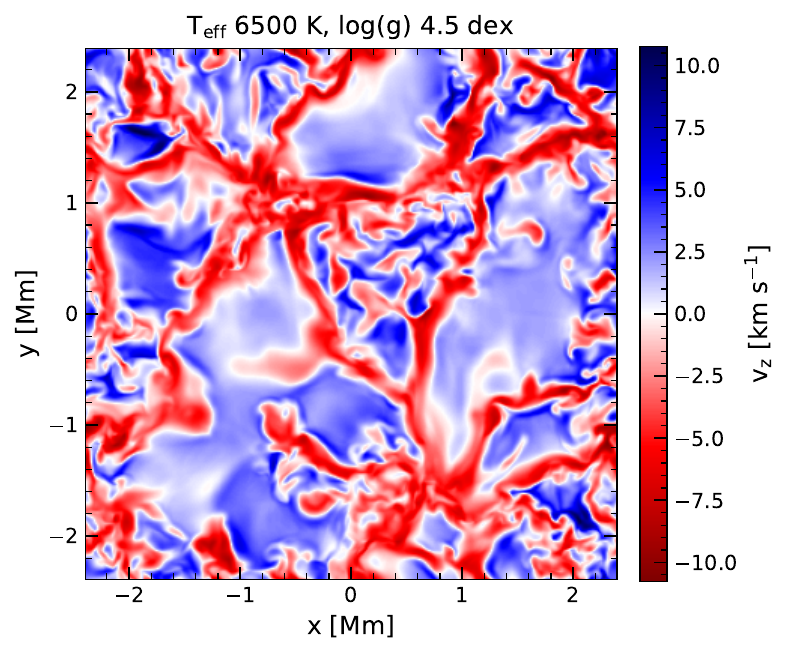}
}
\hbox{
\includegraphics[width=0.33\linewidth]{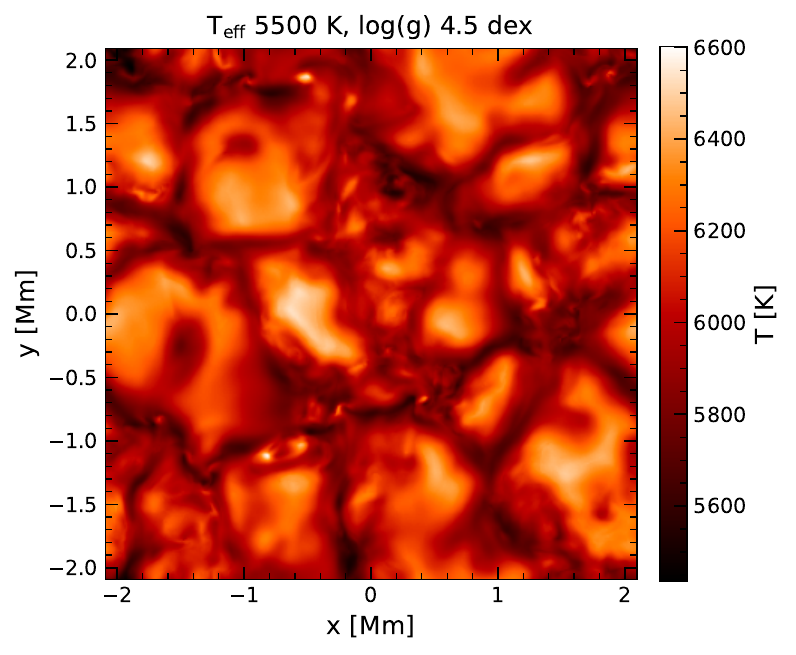}
\includegraphics[width=0.33\linewidth]{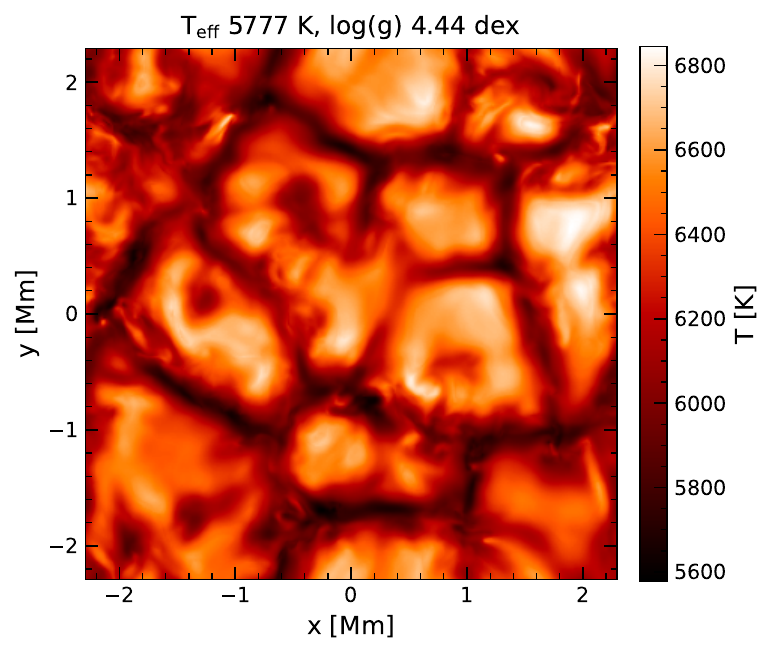} 
\includegraphics[width=0.33\linewidth]{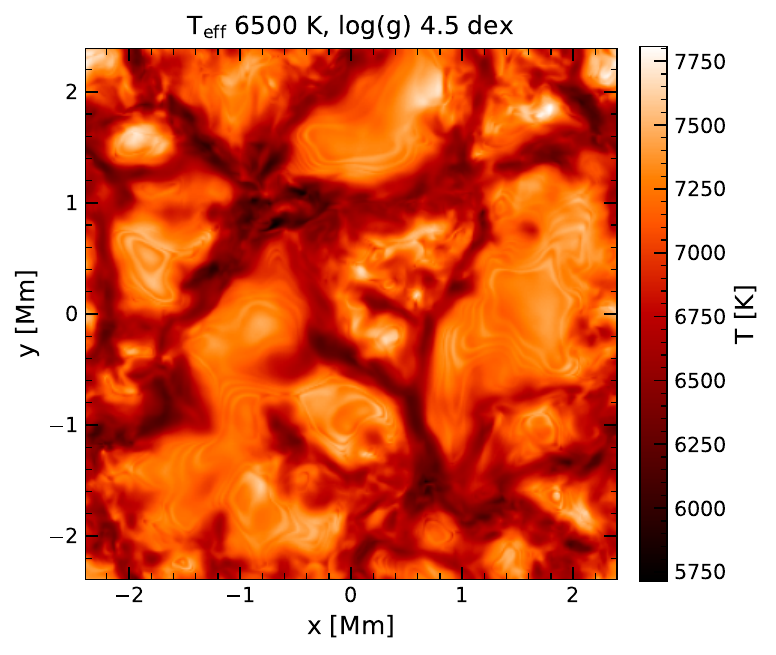}
}
\caption{Vertical component of the velocity field (top) and temperature (bottom) at the optical surface for 3 selected solar-like star simulations including our \texttt{M3DIS} solar model (D).}
\label{fig:uzsurface_and_tcube_other_stars}
\end{figure*}
\subsection{Solar atmospheric structure}
\label{subsec:solar_structure}
In Fig. \ref{fig:model_comparison}, we compare our reference 3D RHD solar model with models created by other groups. Specifically, we compare the mean vertical stratification of our best solar model with the solar \texttt{STAGGER} \citep{Bergemann2012, Magic2013a}, \texttt{CO$^5$BOLD} \citep{Freytag2012, Magg2022}, and the public version of the solar \texttt{MARCS} 1D hydrostatic model\footnote{\url{https://marcs.astro.uu.se}}. The averaging of all 3D cubes is performed on the Rosseland optical depth scale. In the case of \texttt{CO$^5$BOLD} average 3D models are used and hence rms velocities are omitted. For \texttt{STAGGER} only the snapshot resembling the solar effective temperature closest was used.
Overall, the agreement between \texttt{STAGGER}, \texttt{CO$^5$BOLD} and our model is very good. The mean temperature structures (top panel) agree withing $\rm 4\ \%$. Over the entire optical depth range relevant to radiative energy transport \citep{Grupp2004}, the structure of our model agrees to $\rm \sim 2 \%$ with \texttt{STAGGER} model, and below $\rm \sim 1 \%$ with the average \texttt{CO$^5$BOLD} model. The difference with \texttt{CO$^5$BOLD} reaches its maximum below the optical surface at $\rm \log \tau_{ross} \sim 1$ and falls in the sub-percent regime at $\rm \log \tau_{ross} \sim -4$. In general, the \texttt{STAGGER} model is hotter across all optical depths, while the \texttt{MARCS} model transitions from cooler outer layers to a hotter optical surface, and a cooler interior. No systematics can be identified for the \texttt{CO$^5$BOLD} model. As outlined in Sect. \ref{sec:resolution}, the structural differences can primarily be attributed to the resolution of the cooling peak in the region of the steep temperature gradient.  

In the inner layers, conditions closer to the diffusion approximation impose rather narrow range of solutions and hence the structure from different codes converge to similar values. However, in the photospheric layers dominated by radiation loss, differences in the underlying treatment of microphysics, such as the opacity data, abundances, equation-of-state, and binning, can cause systematic differences in the average stratification. Specifically, the slight deviation in the outer temperature structure with the \texttt{STAGGER} model could be caused by the differences in the line blanketing, and specifically in the procedure used for the optimisation of the group opacity bins. In terms of density, the systematic difference of all 3D models with \texttt{MARCS} is the natural consequence of hydrodynamics. Convection provides additional turbulent pressure support to the atmosphere that implies, at a given temperature, lower pressures and densities compared to 1D hydrostatic models (Fig. \ref{fig:T-Pg_3D_MARCS} in Appendix), for a detailed discussion and analysis see \citet{Stein1998} and \citet[][their Fig. 7]{Rosenthal1999}, which is also the mechanisms that changes the properties of the resonant cavity of solar oscillations yielding systematic differences in the properties of oscillation frequencies compared to 1D hydrostatic models \citep[see also e.g.][]{Stein2001, Samadi2003, Houdek2017}. 

The average density stratification, as well as the root-mean-square (rms) velocity distribution, are also in good agreement with \texttt{STAGGER}. On average, the density of our model is 0.02 dex lower, which corresponds to $\rm \sim 5 \%$. Especially at the surface this is in line with the differences in the chemical composition between \cite{Magg2022} and \cite{Asplund2009}, of which the former predicts a higher bulk metallicity but also higher abundances of Fe, Mg, and Si, which are important electron donors at conditions of FGK-type atmospheres. Towards the deeper layers the resolution of the atmosphere again becomes more relevant. The \texttt{MARCS} model shows the same trend, however the differences are more significant and exceed $\rm 0.1\ dex$ below the surface.

Considering that the differences between the RHD models are relatively small we decide that no further refinement, neither of the physical simulation domain nor in the opacity binning, is needed to obtain a realistic atmospheric model on reasonable time-scales. We will further justify this decision in Sect. \ref{subsec:solar_spectrum}, where we test the validity of our solar model by producing observables that can be directly compared to solar observations.
\begin{figure*}
\hbox{
\includegraphics[width=0.33\linewidth]{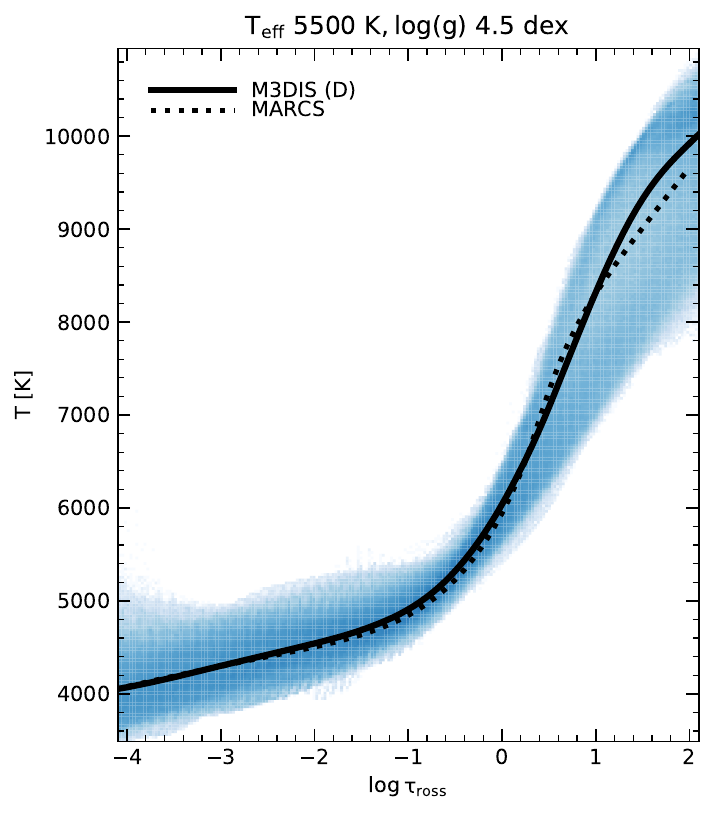}
\includegraphics[width=0.33\linewidth]{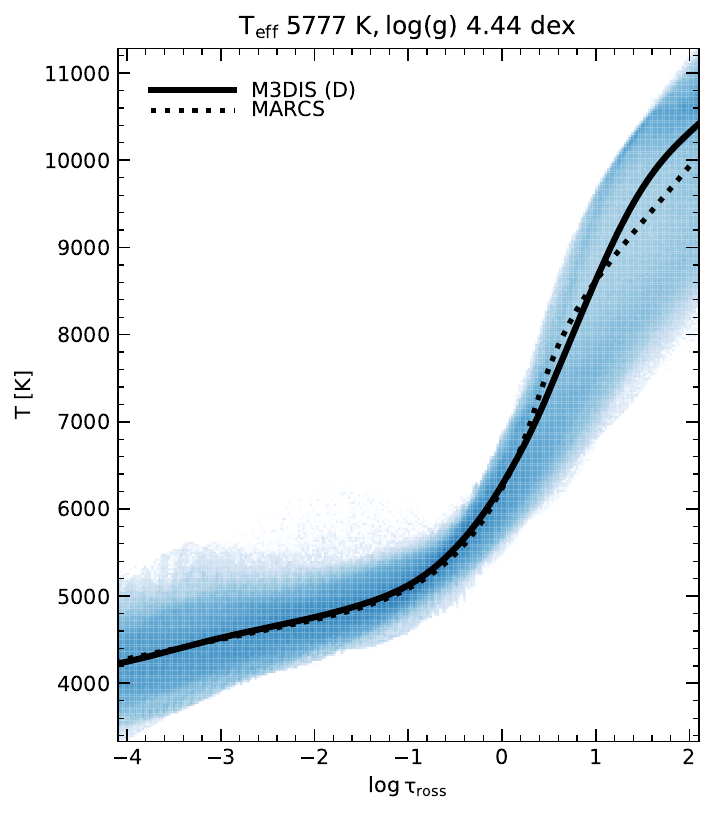} 
\includegraphics[width=0.33\linewidth]{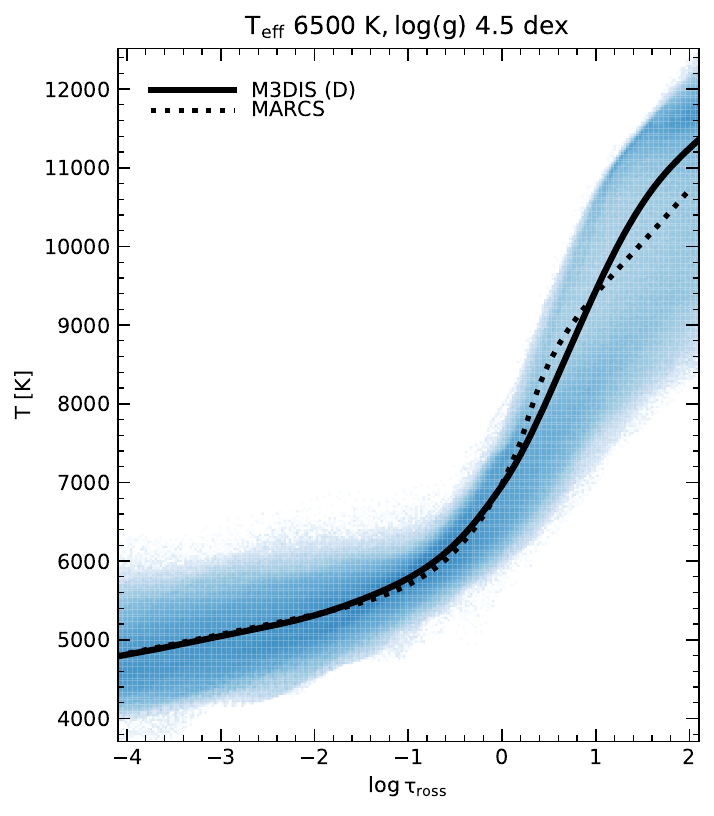}  
}
\caption{Full 3D temperature structure (blue contours) for 3 selected solar-like star simulations including our \texttt{M3DIS} solar model (D). Black solid lines correspond the the average temperature profile, dotted lines show the corresponding \texttt{MARCS} model.}
\label{fig:full3D_temperature}
\end{figure*}
\subsection{Main-sequence models}
We compute several main-sequence 3D RHD models with the same resolution as the solar model D. Specifically, we highlight the models with $\rm \teff = 5500$ and $6500$K, $\logg = 4.5$, and solar metallicity. In Fig. \ref{fig:uzsurface_and_tcube_other_stars} we show horizontal slices of the vertical velocity component (top panels) and temperature (bottom panels) at the optical surface ($\ltross = 0$) in order to illustrate the properties of granulation pattern and its dependence on stellar parameters, specifically on the $\rm \teff$. We also include the solar model for comparison.
The size of the simulation domain is inspired by the corresponding \texttt{STAGGER} model and it is set to be $\rm 4.2,\ 4.6,\ and\ 4.8\ Mm$ in x and y, and $\rm 2.1,\ 2.3,\ and\ 2.4\ Mm$ in z direction, respectively. A representative horizontal extent of the box is important to simulate a sufficiently large number of granules, in order to establish realistic horizontal flows. In other words,  the material in each granule must have enough room to spread sideways and sink back towards the interior, when it reaches the optical surface. If it crosses the periodic border too quickly, the convective pattern can not be established without being significantly influenced by the boundary conditions. \cite{Magic2013a} typically choose approximately ten granules as a satisfactory condition, which is comparable to our results and it suggests that the simulations shown in Fig. \ref{fig:uzsurface_and_tcube_other_stars} are sufficiently large. 

For all models, we furthermore show the full 3D temperature distribution as a function of optical depth in Fig. \ref{fig:full3D_temperature}. We also include the T($\tau$) relationships of their respective horizontal averages and of the corresponding \texttt{MARCS} models with the same stellar parameters. We remind that the MARCS models are in hydrostatic equilibrium, and convective flux transport is represented by the mixing-length theory \citep{Henyey1965} with the mixing-length parameter $\alpha$ set to 1.5  \citep{Gustafsson2008}. A brief summary of the latter models is also provided in \citet{Gerber2023}.

What has been established in Sect. \ref{subsec:solar_structure} can also be observed when investigating 3D RHD models of different $\rm \teff$. The agreement with 1D hydrostatic \texttt{MARCS} models is satisfactory towards to outer (optically thin) regions of the stellar atmospheres. This is simply because in these layers, the structure of solar-metallicity models is close to the radiative equilibrium. We remark that this is not true for metal-poor model atmospheres, because of the prevalence of the adiabatic cooling over radiative heating, as extensively discussed in the literature \citep[e.g.][]{Collet2007, Bergemann2012}. Moving inwards, around $\ltross \sim -2$ to $-1$, the \texttt{MARCS} models are slightly cooler than the average 3D model, and this behaviour inverts in the deeper optically thick layers around $\rm \log \tau_{\rm ross} \gtrsim +1$. At the bottom boundary, the \texttt{MARCS} models are always cooler than their 3D RHD counterparts. For the hottest models in our sample, $\teff = 6500$ K, the difference between the 3D RHD model and the MARCS model amounts to $\sim 500$ K, however, we remark that this difference can in principle be reduced or amplified using a different mixing length parameter \citep[e.g.][]{Fuhrmann1993, Ludwig2009b}. It is also interesting to point out that the extent of temperature inhomogeneities (at a given optical depth) varies significantly with the $\teff$ of the 3D RHD model. As such, the cooler main-sequence model with $\teff = 5500$ K shows a reduced temperature spread at any $\ltross~$compared to the hotter ($\teff = 6500$ K) model, which is due to a more vigorous convection and consequently significantly larger velocity amplitudes in the latter (see also Fig. \ref{fig:uzsurface_and_tcube_other_stars}). 

\begin{figure*}
\centering
\hbox{
\includegraphics[width=0.51\linewidth]{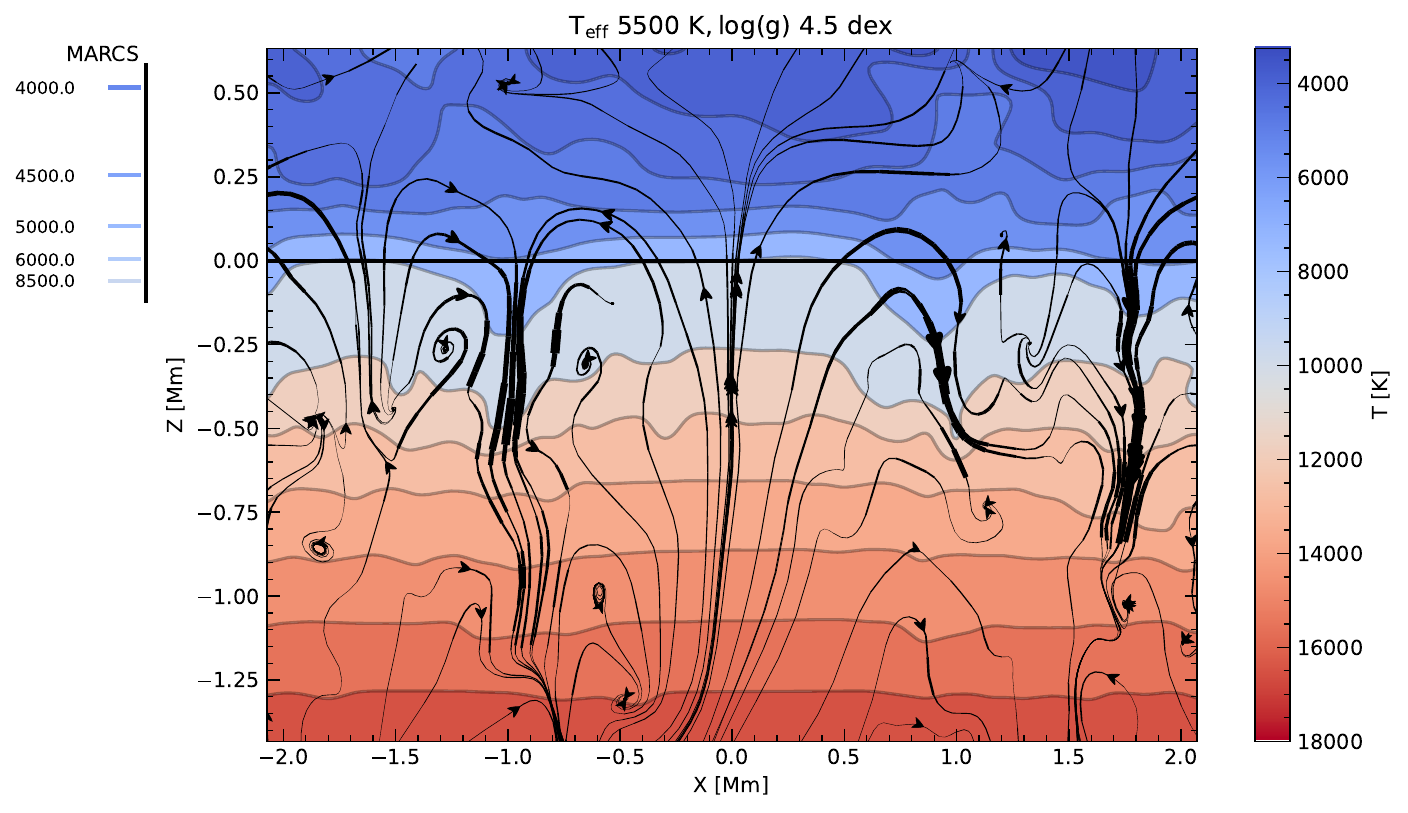} 
\includegraphics[width=0.47\linewidth]{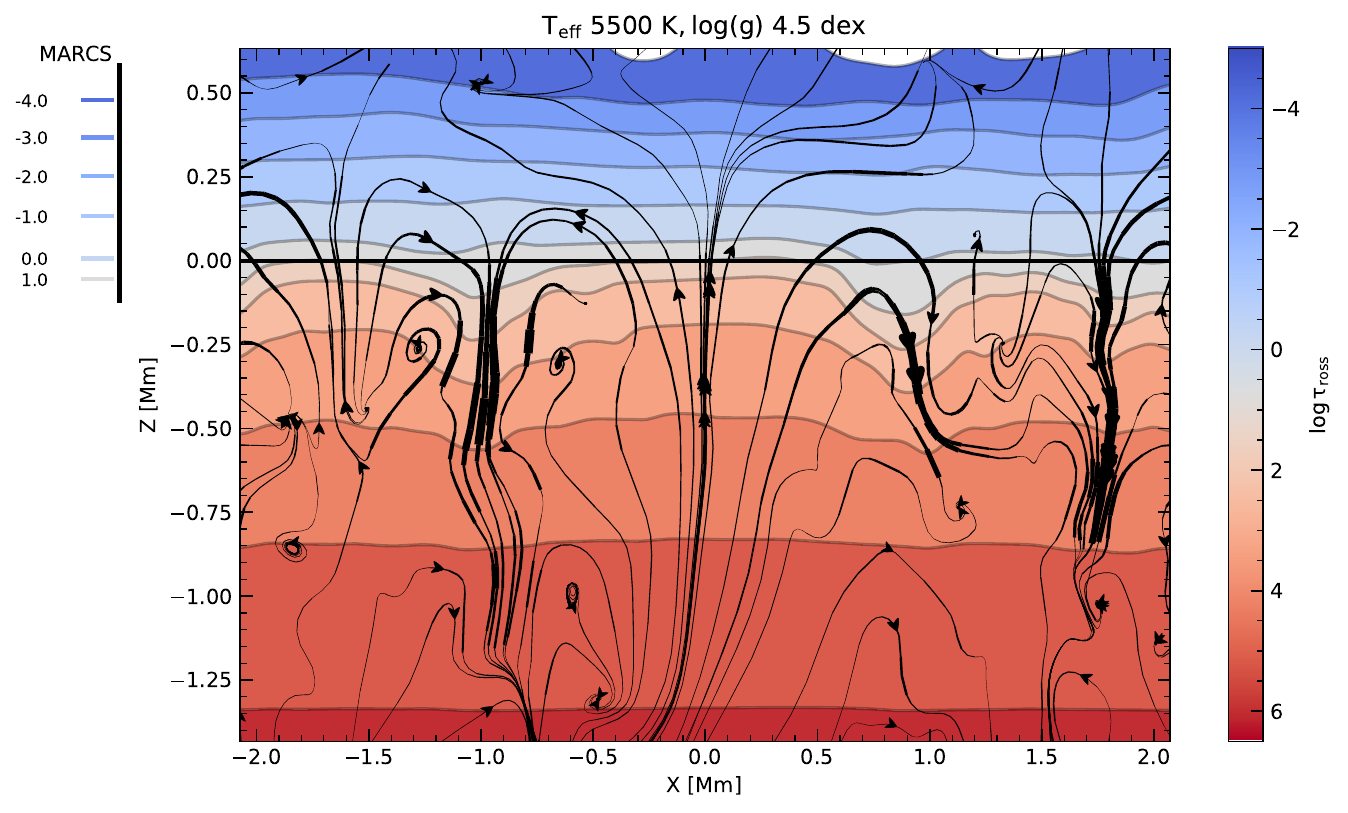}
}
\hbox{
\includegraphics[width=0.51\linewidth]{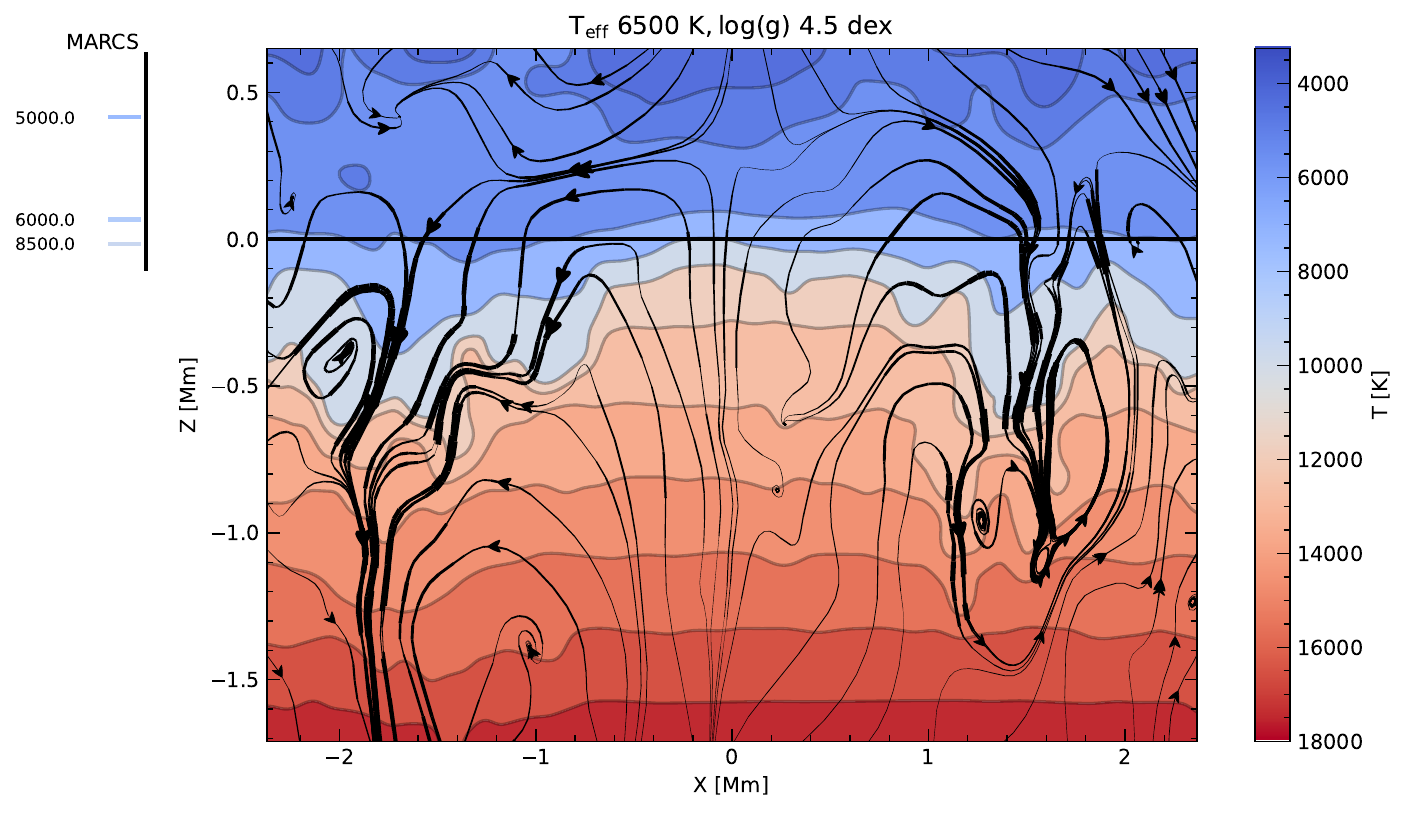} 
\includegraphics[width=0.47\linewidth]{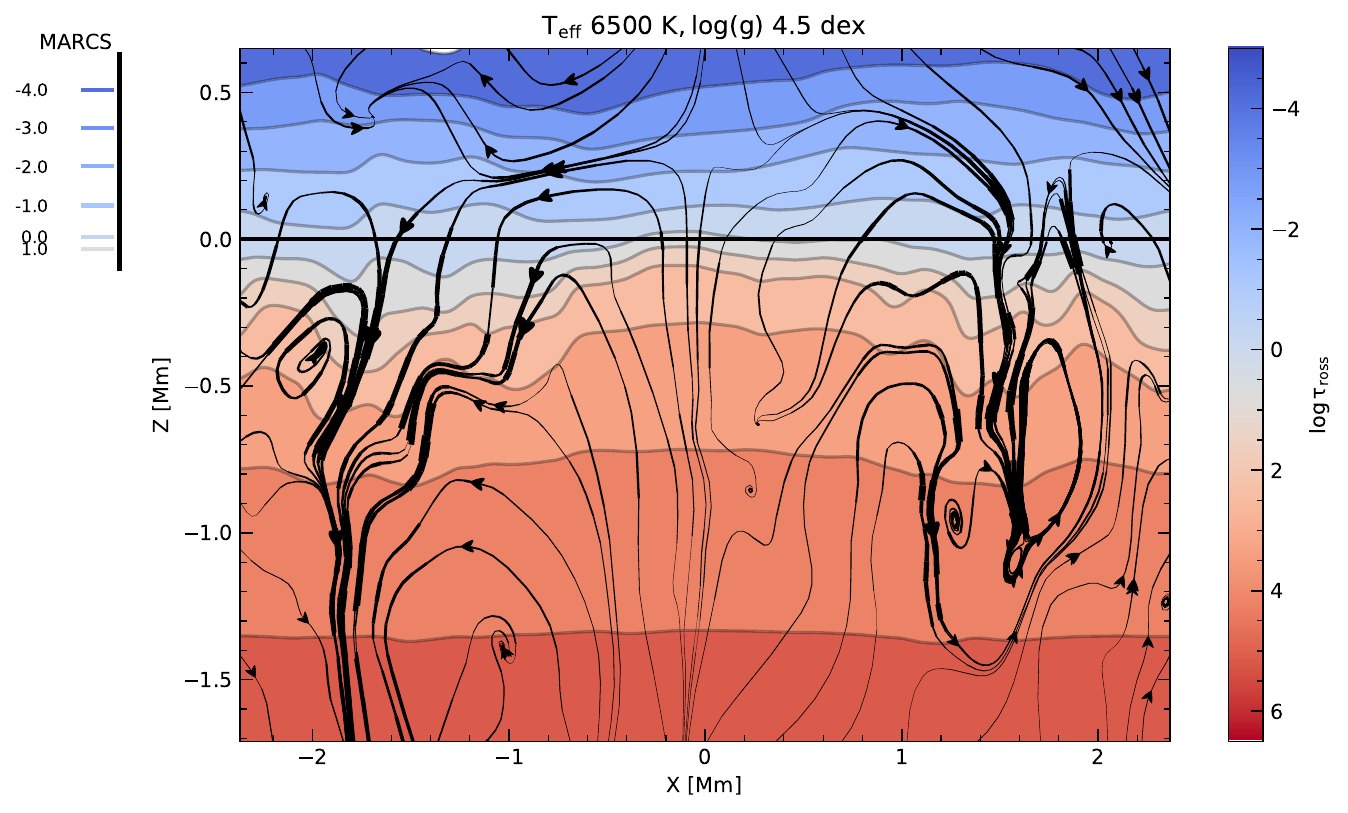} 
}
\caption{Vertical slice through the 3D RHD simulation domain for two selected main-sequence stars (see Tab. \ref{tab:resolution}) with solar metallicity, $\rm \log g$ = 4.5, and $\rm T_{\rm eff} = 5500$ K and $\rm 6500$ K. Note the differences in the vertical sizes of the boxes. Coloured contours corresponds to iso-temperature (left) and iso-$\rm \log \tau_{ross}$ surfaces (right). To the left of each panel, the geometric extension and respective profile of the corresponding 1D HE \texttt{MARCS} model, computed using the same stellar parameters ($\rm T_{\rm eff}$, $\rm \log g$, [Fe/H], and micro-turbulence of 1 kms$^{-1}$) is shown.}
\label{fig:vertical_slice_velocity}
\end{figure*}

Complementary to Fig. \ref{fig:uzsurface_and_tcube_other_stars}, we show vertical slices of the velocity field in Fig. \ref{fig:vertical_slice_velocity} overlaid with coloured contours of iso-temperature (top panels) and iso-optical depth (bottom panels). The direction of the flow is indicated with black streamlines, their width is proportional to the amplitude of velocity. For comparison, we show the exact structure and vertical extent of the equivalent 1D \texttt{MARCS} model to the left of each panel. Levels of temperature and $\rm \log \tau_{ross}$, as well as the vertical black line showing the extent of the 1D HE models, are placed at their respective geometrical height in comparison to the 3D model.

Firstly, it should be noted that the 1D HE models are much shallower than 3D RHD models that simulate convection and turbulent motions from first principle. Secondly, as already seen in 2D projections (Fig. \ref{fig:full3D_temperature}), while the average structures are not too inconsistent between 1D HE and 3D RHD, the crucial difference between the two types of model atmospheres is in the presence of substantial horizontal inhomogeneities in all thermodynamic quantities (T, $\rho$, $P_{\rm gas}$, $\ltross$, etc.) associated with horizontal and vertical mass motions. In all 3D models, multiple regions of pronounced downward motions are observed, with some extending across the entire vertical simulation domain. In between these regions, there are upflows of smaller vertical amplitude, however typically much greater horizontal extent. These upflows pivot near the optical surface, which leads to the familiar granular pattern observed at the surface \citep[e.g.][]{Stein1998, Nordlund2009}. 

Fig. \ref{fig:vertical_slice_velocity} clearly shows that the flow pattern disrupts the iso-$\rm T$ and $\ltross$~surfaces from the simple linear stratification seen in hydrostatic models, and these undulations in T and opacity are highly correlated. The curvature is especially pronounced in the hotter models (bottom panels in Fig. \ref{fig:vertical_slice_velocity}). The updrafts pull the hotter material along, bringing it closer to the surface, while the downdrafts mix cooler surface material deeper within the star. This has significant implications for the opacity and consequently for the optical depth. Specifically, in the intergranular lane the same optical depth is achieved at much larger geometric depths in the atmosphere, compared to the granule itself. In other words, one can see deeper into the intergranular lanes than into the granules. For the 6500 K model, the geometric extent (vertically) of the iso-$\ltross$~surface may exceed 400 km. By analogy, the vertical location of the iso-thermal surface changes by the same amount depending on the horizontal position.  Also in the optically thin layers, a very large difference between the temperature of the granules and of the intergranular lanes is observed, leading to an approximately $1\,000$ K difference between these two main morphological components of the velocity field. Overall, the intergranular velocity field, its downward extent, and influence on the atmospheric structure become more significant with increasing effective temperature and/or decreasing the surface gravity of the models \citep{Magic2013a}.
\subsection{Solar spectrum: validation and examples} \label{subsec:solar_spectrum}
\begin{figure*}
\centering
\hbox{
\includegraphics[width=0.5\linewidth]{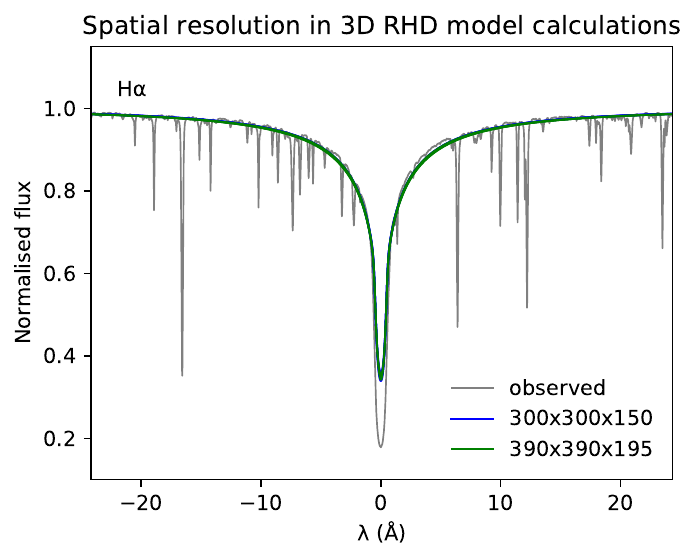}
\includegraphics[width=0.5\linewidth]{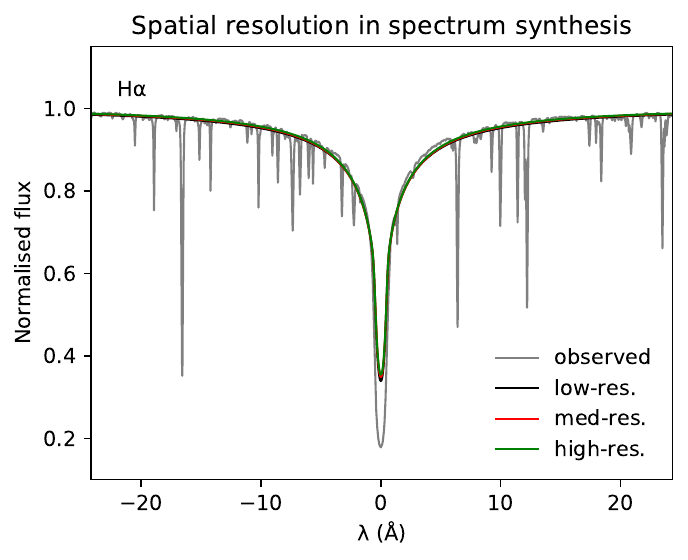}
}
\hbox{
\includegraphics[width=0.5\linewidth]{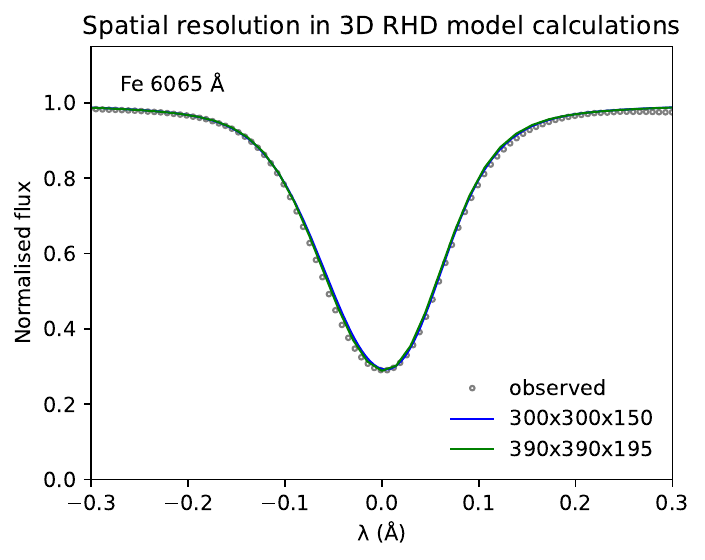}
\includegraphics[width=0.5\linewidth]{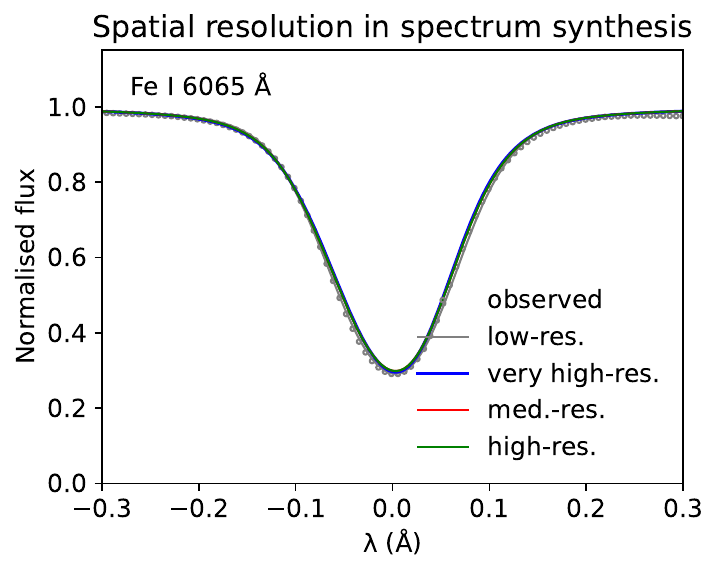}
}
\caption{Comparison of $\ha$ and Fe I line profiles computed with \texttt{M3DIS}, but with different spatial resolutions used in 3D RHD model calculations (left column) and in the post-processing spectrum synthesis (right column). For the results presented on the left-hand side, we adopt models C and D from Tab. \ref{tab:resolution}, respectively, where D $(\rm N_x \times N_y \times N_z) = (300,300, 150)$ is our reference. For the results presented on the right-hand side, we use 3 characteristic spatial resolutions that differ only in horizontal resolution of $(\rm N_x \times N_y)= (10,10)\ \rm{ (low-res.)},\ (30,30)\ \rm{ (med.-res.)},\ and\ (80,80)\ \rm{ (high-res.)}$. For Fe I, we also show for comparison the profile computed a very high horizontal resolution of $(150,150)$ in spectral synthesis. }See text.
\label{fig:spec}
\end{figure*} 
The main aim of our work is to develop a self-consistent framework to compute model atmospheres and spectrum synthesis with detailed line formation in NLTE, with as far as possible consistent micro-physics. In this section, we present the analysis of synthetic line profiles for the Sun, computed using two approaches. In one approach, we test the influence of spatial resolution in 3D RHD simulations (Sect. \ref{sec:resolution}) on the properties of spectral lines. This has already been explored earlier in the context of atmospheric structure. In the second approach, we test the influence of the spatial resolution in the post-processing spectrum synthesis, which is carried out as described in Sect. \ref{subsec:spectrum_synthesis}. Two tests are performed, in order to assess the needs in terms of precision of the thermodynamic structure of the \texttt{M3DIS} 3D RHD model, as well as in terms of the spatial resolution in the post-processing spectrum synthesis calculations. The latter aspect was also addressed in \citet{Bergemann2019} for Mn I and in \citet{Bergemann2021} for O I lines. We use LTE line formation, because it provides a more stringent test of the sensitivity of line formation to the local underlying atmospheric structure. In all cases presented here, the radiation transfer is computed in full 3D.

\begin{figure*}
\includegraphics[width=1.0\linewidth]{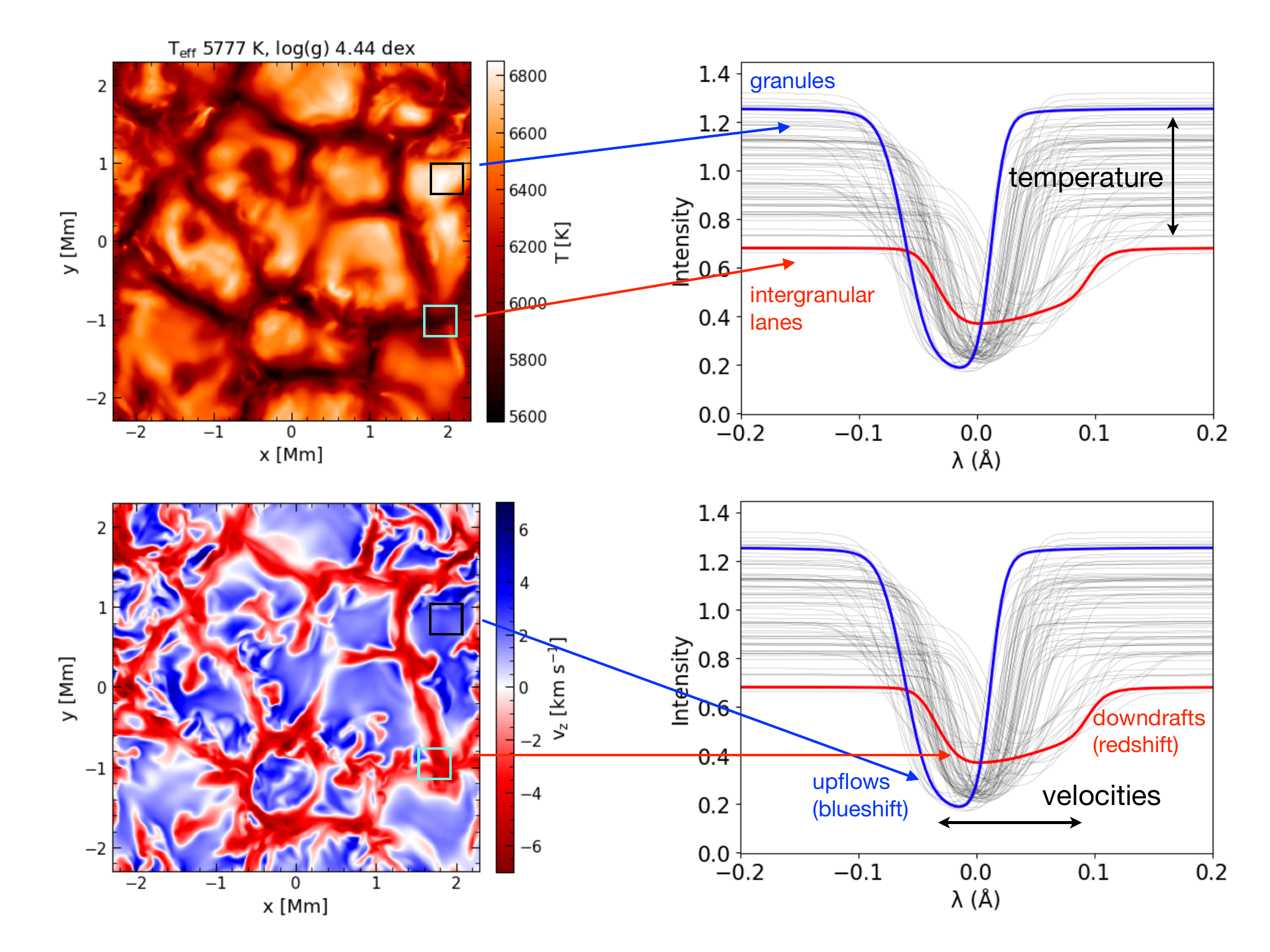}
\caption{Simplified diagram of line formation in inhomogeneous atmospheres. Left panels: Temperature (top) and vertical velocity (bottom) slices at the optical surface. Right panels: Fe I 5432 \AA~ line intensity emerging from different locations on the stellar surface.}
\label{fig:demo}
\end{figure*}

In Fig. \ref{fig:demo} we show a simplified diagram of line formation in inhomogeneous atmospheres. This diagram helps to visualise the connection between the structural properties of 3D models, specifically, temperature, density and convective velocity fields, and the observed quantities, that is, the emergent spectrum. We illustrate how the Fe I line at 5432 \AA~vary across the simulation cube and identify graphically the main features that follow from the complex temperature distribution and form the complex velocity structure within granules and intergranular lanes.

\subsubsection{Linelist}
For these two tests, we use the hydrogen Balmer $\ha$ line and a set of diagnostic Fe I lines. The Fe I lines were carefully chosen from the atomic data compilations published by the University of Wisconsin group\footnote{\url{https://github.com/vmplacco/linemake/tree/master/mooglists}}, VALD database\footnote{\url{http://vald.astro.uu.se}}, and the Gaia-ESO linelist \citep{Heiter2021}. The line selection procedure was iterative and involved manual inspection of the line profiles in the observed solar spectra (to avoid blended features), lines with reliable $\log gf$ values (as for example given by the Y flag in the Gaia-ESO compilation), and relevance for diagnostics of the Sun (based on \citealt{Magg2022}) and for metal-poor stars as gauged by the availability in the University of Wisconsin compilation. We also aim to sample a range of strengths, excitation potentials and wavelengths from $\rm 5000$ to $\rm 9000$\,\AA, the range that is used for optical spectroscopic surveys. With all these selection criteria, our final list includes 22 Fe I lines. Their parameters are provided in Table \ref{tab:lines}.

\subsubsection{Influence of 3D RHD model resolution}
\begin{figure}
\includegraphics[width=1.0\columnwidth]{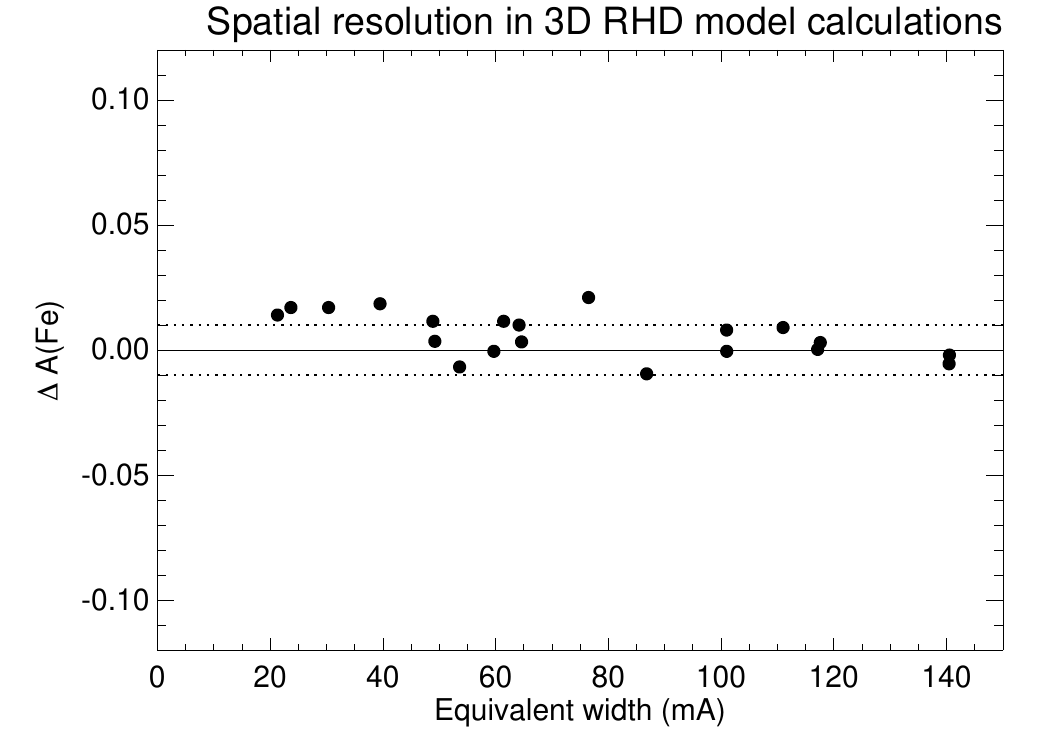}
\caption{Error incurred in the Fe abundance as a function of equivalent line width corresponding to the difference between the high and very-high resolution models D and C, respectively. The horizontal lines show the 0.01 dex mark, to guide the eye. The ranges of y-axes are chosen deliberately to illustrate the typical error in abundances caused by the error of atomic data that, for diagnostic solar lines, ranges from 0.03 to 0.1 dex \citep{Magg2022}.}
\label{fig:fe_md_hd}
\end{figure}

In the left panels of Fig. \ref{fig:spec} we compare the $\ha$ and Fe I line profiles calculated using our reference \texttt{M3DIS} solar model D (Tab \ref{tab:resolution}) with the profiles computed using the model with the highest vertical resolution of 5 km (C) that corresponds to (x,y)$\rm =(390, 390, 390)$. We stress that in this test (in contrast to the 2nd test performed below), the model structure is different, whereas the resolution of spectrum synthesis is unchanged. The profiles correspond to averages over four snapshots taken at time intervals of $100$ seconds. The agreement of the line profiles is very good, especially for $\ha$, which has been traditionally used \citep[e.g.][]{Barklem2002, Mashonkina2008, Ruchti2013} as a diagnostic of atmospheric structure of FGK-type stars and consequently represents an important $\rm T_{\rm eff}$ indicator. For $\hb$, we find small differences in the line wings, corresponding to the difference of the order $\rm 20$ to $\rm 50$ K. Also for the Fe I lines, the agreement between LTE profiles computed at different resolutions in the RHD calculations is excellent. This is further illustrated in Fig. \ref{fig:fe_md_hd}, where we show the error incurred in the Fe abundance corresponding to the difference in structure between the high and very-high resolution 3D RHD models. For most Fe I lines, the error is within $\rm 0.01$ dex and there is no systematic trend with the line equivalent width. The lines with a slightly larger (positive) error of $\rm +0.015$ dex are all near-IR lines with wavelengths $\rm \gtrsim 8000$ \AA, whereas the features with the negative error are those towards the blue, at $\rm \lambda \sim 5000$ \AA. Overall, however, the impact of the resolution in radiation hydrodynamics calculations appears to be modest, which suggests that our standard geometric setup (corresponding to that of model C in Tab. \ref{tab:resolution}) can be reliably used for modelling main-sequence FGK-type stars stars.  

\subsubsection{Influence of spatial resolution in spectrum synthesis}
In the second test, we compare and test different spatial resolutions used in the post-processing spectrum synthesis, while using the same underlying 3D RHD model. This test is valuable, because the motivation behind our work is the computationally expensive solution of the NLTE problem. In this context, decreasing the horizontal resolution during the calculations of departure coefficients offers a huge computational speed-up compared to solving coupled RT and statistical equilibrium equations at each iteration. Therefore, it is important to us to find a balance between the computational demands and complexity of the approach. Such a horizontal down-sampling is among the most powerful means to achieve the speed-up and we explore this approach in this section.

The model is re-sampled on the geometrical scale using linear interpolation. The number of points in the vertical is chosen to be 299, which is exactly twice the number of grid points in our best solar model (D), minus 1 point (see Tab. \ref{tab:resolution}). This ensures that every second grid node coincides with the original grid of the snapshot to minimise interpolation errors, which is more relevant in the vertical direction. We find that the effect on spectral lines is small. In Fig. \ref{fig:spec} (right panel) we illustrate the effect for the $\ha$ line, as well as for one of the optical Fe I lines at 6065 \AA. The comparison is performed for three models, each down-sampled to an effective spatial resolution of (x,y,z) $\rm = (10,10,299)$, $\rm (30,30,299)$, and $\rm (80,80,299)$, respectively. Each profile corresponds to the average of a series of nine snapshots. 

As seen in the right panels of Fig. \ref{fig:spec}, for $\ha$, the synthetic profiles remain very close as long as the horizontal resolution in the spectrum synthesis calculations exceeds (x,y)$\rm =(10,10)$. This is in agreement with our earlier studies \citep{Bergemann2019}, where we found a very small effect on the profiles of Mn lines. For Fe I, we also show the effect of very high horizontal resolution, that is, $\rm(x, y)=(150,150)$. Here, we find that the profiles computed with the model at $\rm (x, y)=(10,10)$ are slightly stronger compared to the profiles computed at $\rm (x, y)=(30,30)$, $\rm (x, y)=(80,80)$, and $\rm (x, y)=(150,150)$, whereas the latter three simulations are in excellent agreement. Comparing other line profiles, we find that the effect of low spatial resolution in spectrum synthesis slightly over-estimates the strength of line profiles and thus may under-estimate the Fe abundance by $\rm \sim 0.02$ to $\rm 0.05$ dex. This is a typical uncertainty that shall be accounted for in abundances inferred through fits of LTE synthetic spectra, when low horizontal resolution is employed. Such tests are vastly more time-consuming and require substantially more computational resources in NLTE, however, in NLTE the effects are expected to be smaller. This is because the effects of differences in temperature and density are usually amplified in LTE calculations, due to the direct coupling of opacity and the source function to the local value of temperature. In contrast, in NLTE line opacity and source function are strongly influenced by the background non-local radiation field, especially in the blue and near-UV regime, that forms deeper in the atmosphere of FGK-type stars. This decoupling effectively eliminates the strong temperature dependence of the source function \citep{Bergemann2012}.
While it may appear surprising at first that an under-sampling where horizontal structures are certainly severely degraded still can results in accurate results, it should be realised that on the other hand there is in some sense an over-representation of structures, with several granules providing samples at any one time, and with several (here nine) samples in time.

\begin{figure*}
\centering
\includegraphics[width=0.91\linewidth]{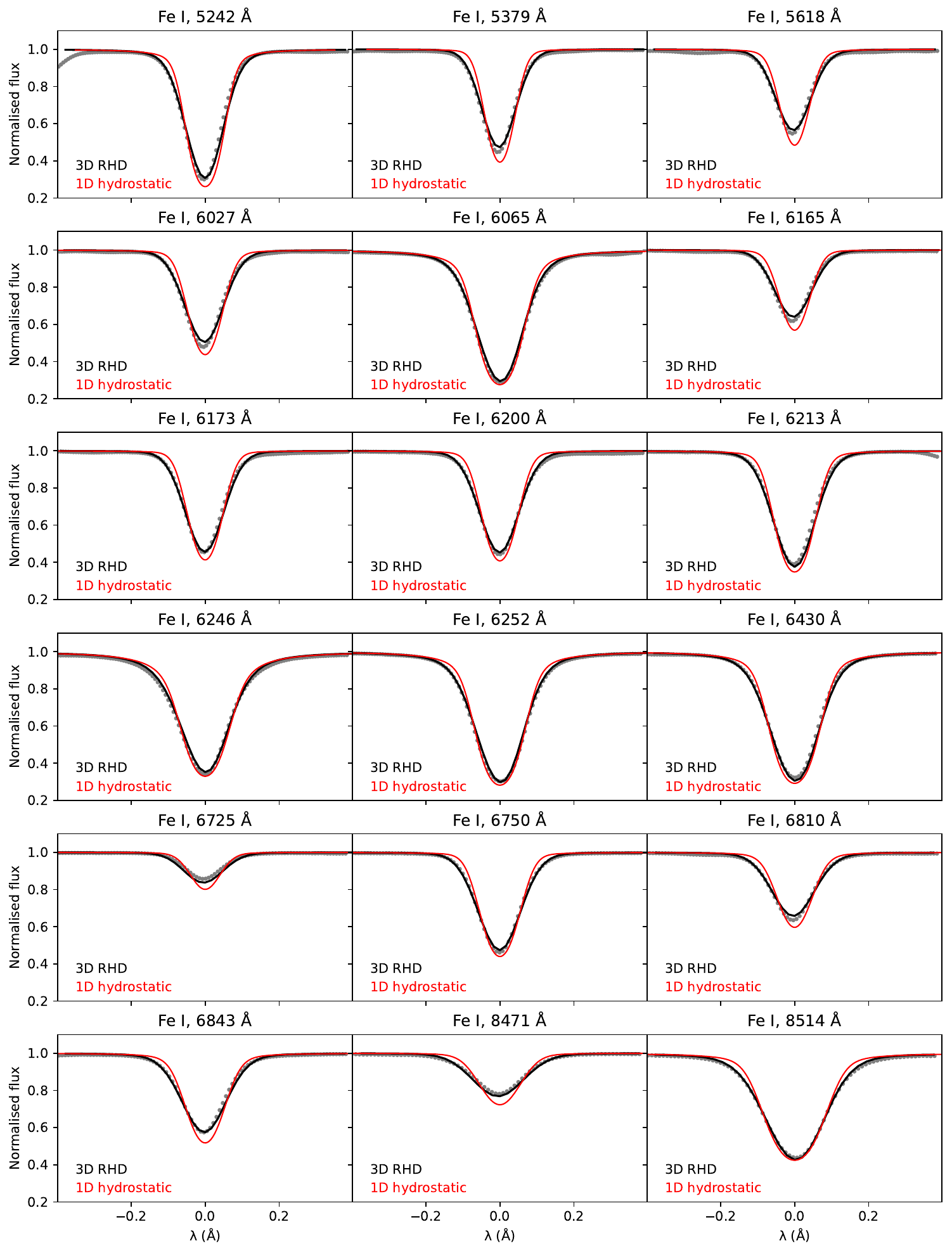}
\caption{Comparison of the observed solar Fe I lines from the FTS KPNO flux atlas \citep{Kurucz1984} with the synthetic line profiles computed using our 3D RHD Dispatch model (3D RHD) and the MARCS solar model (1D hydrostatic). For both models, the same iron abundance of A(Fe)$=7.50$ was used. To compute the MARCS line profiles, we additionally include a microturbulence of 1 kms$^{-1}$. Except the solar rotation, no other broadening was applied and the synthetic lines were not fit to the observations in any way.}
\label{fig:feall}
\end{figure*}

Finally, in Fig. \ref{fig:feall} we show the final line profiles of the Fe I lines compared to the very high-resolution solar flux observations from the KPNO FTS atlas \citep{Kurucz1984}. We overlay the results obtained from the radiation transfer calculations with our high-resolution reference Dispatch 3D RHD solar model (Table 2) and from the MARCS 1D HE solar model. In both cases, we use the same Fe abundance of A(Fe) $=7.50$ and apply a broadening of 1.6 kms$^{-1}$ to account for the solar rotation. In the 1D HE case, we also include the standard microturbulence of 1 kms$^{-1}$. No other velocities have been included, as we want to avoid arbitrary fine-tuning of the models. Clearly, the 3D RHD modelling leads a much superior description of the asymmetric shapes of the observed line profiles compared to the 1D results. The latter, in virtue of the underlying assumptions of micro- and macro-turbulence, always predict perfectly symmetric profiles (note this is distinct from the effects of isotopic structure or hyperfine splitting), thus failing to explain the deeper wings and shallower line cores. Also the 3D models successfully account for the characteristic blueshifts of the lines. This reinforces the evidence that 3D radiation-transfer modelling is paramount for accurate and precise stellar spectroscopy in order to obtain realistic synthetic stellar fluxes and intensities, and consequently to derive accurate fundamental stellar parameters by comparing models to observations. 

%
%
%
%
%
%
\section{Conclusions} \label{sec:discussion}

In this work, we present a new framework to solve radiation-hydrodynamics and detailed radiation transfer in LTE and NLTE, in order to model sub-surface stellar convection and calculate synthetic observables for FGK-type stars. We build upon the Copenhagen legacy RHD code, also known as the \texttt{STAGGER} code, the \texttt{MULTI3D} NLTE radiation transfer code, and the \dis high-performance computing framework \citep{Nordlund2018}. Here we outline the physical basis of the code, the key numerical properties, and updates to the input micro-physics, and illustrate the gain in computational performance of the code, applying it to the simulations of sub-surface convection of the Sun and several stars on the main-sequence. Specifically, we show that with the new \texttt{DISPATCH} framework 3D RHD models with geometric and physical properties suitable for stellar spectroscopic modelling now can be computed -- from scratch -- on timescales of only a few thousand CPU hours, doable on medium-size computer clusters. Also very high resolution calculations, for example with vertical resolution of up to 6 km throughout the entire simulation domain, can be carried out conveniently with much more modest computations, compared to other available codes, permitting detailed systematic analysis of physical properties of the models at much smaller length scales.

We demonstrate how the updates to the code in terms of microphysics influence the basic observables for the Sun, specifically the bulk spectral energy distribution and the centre-to-limb variation for the continuum, by comparing the model predictions to the data from \citet{Neckel1994}. We show that similarly to other models, such as \texttt{STAGGER} and \texttt{CO$^5$BOLD}, our new models reproduce well the solar observations, with a minor difference in the blue and near-UV regime, around $\rm \lambda \lesssim 400$ nm. This difference manifests itself in somewhat steeper centre-to-limb intensity variations, but is common to other 3D RHD models \citep{Beeck2012,Pereira2013}. For the optical and near-IR, the CLV is in excellent agreement with the solar data.

Further, we perform an analysis of the influence of the geometric resolution in 3D radiation-hydrodynamic simulations and in the post-processing spectrum synthesis. We show that low- and medium-resolution models are hotter at the surface, and have slightly lower densities and velocities in the sub-photospheric layers. However, models with high- and very-high resolution show a high degree of consistency in their bulk thermodynamic properties, which results in very similar synthetic observables. For our purposes, which primarily aim to develop a method of characterisation of stellar observations (such as spectra and photometry) the spatial resolution of $(\rm N_x \times N_y \times N_z) = (300,300,150)$ (in hydrodynamics, but double vertical resolution in radiation transfer) is appropriate to achieve the desired precision of the 3D structure.

In the atmospheric layers above $\rm \log \tau_{Ross} \lesssim -3$, the vertical resolution has a systematic effect of decreasing the surface temperature, which may impact the cores of strong lines probing the outermost structure, such as $\ha$, and may bias the spectroscopic results if these are modelled in LTE. We show, however, that the corresponding effect on abundance diagnostics does not exceed $\rm 0.01$ to $\rm 0.02$ dex for the typical diagnostic lines with equivalent widths from 20 to 140\,m\AA. We also find that---as a result of sampling several multi-structure snapshots in time---a spectrum synthesis horizontal resolution of $(\rm N_x \times N_y)$  $= (30,30)$ is fully sufficient for precision diagnostics, as the corresponding error is smaller than the standard uncertainty of the input atomic and molecular data. We expect that even lower resolutions would be suitable for NLTE calculations, and encourage further studies of this kind.

This is the first paper of the series that outlines the methodology, input physics, the properties of the simulations, and presents basic validation tests. In the subsequent papers, we aim to expand the parameter space of the simulations, probing the domain of more evolved stars, but also cooler stars on the main-sequence, to capture the entire parameter space of FGKM type stars that are in the prime focus of upcoming large-scale stellar surveys, specifically the 4MOST high-resolution survey of the Galactic disc and bulge \citep{deJong2019,Bensby2019}, the SDSS-V Milky Way Mapper \citep{Kollmeier2017, Straumit2022}, and PLATO \citep{Rauer2014, Gent2022}. As we develop the approach and methodology, we will put a substantial effort into developing an easily-accessible and secure repository for public access, building upon our experience with 1D HE model atmospheres and departure grids \citep{Gerber2023} and NLTE abundance corrections\footnote{\url{nlte.mpia.de}}. \texttt{M3DIS} stellar models will thus provide a qualitatively new basis for accurate and precise diagnostics of fundamental stellar parameters and detailed chemical composition of stars. The version of \texttt{DISPATCH}, on which the present code is built, is not publicly available yet. However, there are plans to also release the code in the near future.

\section*{Acknowledgements}

We thank the referee for the careful review and many insightful comments that have helped to significantly improve the paper. MB is supported through the Lise Meitner grant from the Max Planck Society. PE is supported through the IMPRS fellowship of the Max Planck Society.
This project has received funding from the European Research Council (ERC) under the European Unions Horizon 2020 research and innovation programme (Grant agreement No. 949173). We acknowledge support from the DAAD grant Project No.: 57654415. Computations were performed on the HPC system Raven and Cobra at the Max Planck Computing and Data Facility (MPCDF). 
 \bibliographystyle{aa}
 \bibliography{aanda.bbl} 

\begin{thebibliography}{97}
\expandafter\ifx\csname natexlab\endcsname\relax\def\natexlab#1{#1}\fi

\bibitem[{{Ali} \& {Griem}(1965)}]{AliGriem1965}
{Ali}, A.~W. \& {Griem}, H.~R. 1965, Physical Review, 140, 1044

\bibitem[{{Asplund} {et~al.}(2009){Asplund}, {Grevesse}, {Sauval}, \&
  {Scott}}]{Asplund2009}
{Asplund}, M., {Grevesse}, N., {Sauval}, A.~J., \& {Scott}, P. 2009, \araa, 47,
  481

\bibitem[{{Barklem} \& {Piskunov}(2003)}]{BarklemPiskunov2003}
{Barklem}, P.~S. \& {Piskunov}, N. 2003, in Modelling of Stellar Atmospheres,
  ed. N.~{Piskunov}, W.~W. {Weiss}, \& D.~F. {Gray}, Vol. 210, E28

\bibitem[{{Barklem} {et~al.}(2000){Barklem}, {Piskunov}, \& {O'Mara}}]{BPO2000}
{Barklem}, P.~S., {Piskunov}, N., \& {O'Mara}, B.~J. 2000, \aap, 363, 1091

\bibitem[{{Barklem} {et~al.}(2002){Barklem}, {Stempels}, {Allende Prieto},
  {Kochukhov}, {Piskunov}, \& {O'Mara}}]{Barklem2002}
{Barklem}, P.~S., {Stempels}, H.~C., {Allende Prieto}, C., {et~al.} 2002, \aap,
  385, 951

\bibitem[{{Beeck} {et~al.}(2012){Beeck}, {Collet}, {Steffen}, {Asplund},
  {Cameron}, {Freytag}, {Hayek}, {Ludwig}, \& {Sch{\"u}ssler}}]{Beeck2012}
{Beeck}, B., {Collet}, R., {Steffen}, M., {et~al.} 2012, \aap, 539, A121

\bibitem[{{Bell} \& {Berrington}(1987)}]{Bell1987}
{Bell}, K.~L. \& {Berrington}, K.~A. 1987, Journal of Physics B Atomic
  Molecular Physics, 20, 801

\bibitem[{{Bensby} {et~al.}(2019){Bensby}, {Bergemann}, {Rybizki}, {Lemasle},
  {Howes}, {Kovalev}, {Agertz}, {Asplund}, {Barklem}, {Battistini},
  {Casagrande}, {Chiappini}, {Church}, {Feltzing}, {Ford}, {Gerhard},
  {Kushniruk}, {Kordopatis}, {Lind}, {Minchev}, {McMillan}, {Rix}, {Ryde}, \&
  {Traven}}]{Bensby2019}
{Bensby}, T., {Bergemann}, M., {Rybizki}, J., {et~al.} 2019, The Messenger,
  175, 35

\bibitem[{{Bergemann} {et~al.}(2019){Bergemann}, {Gallagher}, {Eitner},
  {Bautista}, {Collet}, {Yakovleva}, {Mayriedl}, {Plez}, {Carlsson},
  {Leenaarts}, {Belyaev}, \& {Hansen}}]{Bergemann2019}
{Bergemann}, M., {Gallagher}, A.~J., {Eitner}, P., {et~al.} 2019, \aap, 631,
  A80

\bibitem[{{Bergemann} {et~al.}(2021){Bergemann}, {Hoppe}, {Semenova},
  {Carlsson}, {Yakovleva}, {Voronov}, {Bautista}, {Nemer}, {Belyaev},
  {Leenaarts}, {Mashonkina}, {Reiners}, \& {Ellwarth}}]{Bergemann2021}
{Bergemann}, M., {Hoppe}, R., {Semenova}, E., {et~al.} 2021, \mnras, 508, 2236

\bibitem[{{Bergemann} {et~al.}(2012){Bergemann}, {Lind}, {Collet}, {Magic}, \&
  {Asplund}}]{Bergemann2012}
{Bergemann}, M., {Lind}, K., {Collet}, R., {Magic}, Z., \& {Asplund}, M. 2012,
  \mnras, 427, 27

\bibitem[{{B{\"o}hm-Vitense}(1958)}]{Boehm1958}
{B{\"o}hm-Vitense}, E. 1958, \zap, 46, 108

\bibitem[{{Bonifacio} {et~al.}(2018){Bonifacio}, {Caffau}, {Ludwig}, {Steffen},
  {Castelli}, {Gallagher}, {Ku{\v{c}}inskas}, {Prakapavi{\v{c}}ius}, {Cayrel},
  {Freytag}, {Plez}, \& {Homeier}}]{Bonifacio2018}
{Bonifacio}, P., {Caffau}, E., {Ludwig}, H.~G., {et~al.} 2018, \aap, 611, A68

\bibitem[{{Caffau} {et~al.}(2011{\natexlab{a}}){Caffau}, {Bonifacio},
  {Fran{\c{c}}ois}, {Sbordone}, {Monaco}, {Spite}, {Spite}, {Ludwig}, {Cayrel},
  {Zaggia}, {Hammer}, {Randich}, {Molaro}, \& {Hill}}]{Caffau2011b}
{Caffau}, E., {Bonifacio}, P., {Fran{\c{c}}ois}, P., {et~al.}
  2011{\natexlab{a}}, \nat, 477, 67

\bibitem[{{Caffau} {et~al.}(2011{\natexlab{b}}){Caffau}, {Ludwig}, {Steffen},
  {Freytag}, \& {Bonifacio}}]{Caffau2011a}
{Caffau}, E., {Ludwig}, H.~G., {Steffen}, M., {Freytag}, B., \& {Bonifacio}, P.
  2011{\natexlab{b}}, \solphys, 268, 255

\bibitem[{{Canuto} \& {Mazzitelli}(1991)}]{Canuto1991}
{Canuto}, V.~M. \& {Mazzitelli}, I. 1991, \apj, 370, 295

\bibitem[{{Carlsson} {et~al.}(2019){Carlsson}, {De Pontieu}, \&
  {Hansteen}}]{Carlsson2019}
{Carlsson}, M., {De Pontieu}, B., \& {Hansteen}, V.~H. 2019, \araa, 57, 189

\bibitem[{{Chandrasekhar}(1939)}]{Chandrasekhar1939}
{Chandrasekhar}, S. 1939, {An introduction to the study of stellar structure}

\bibitem[{{Collet} {et~al.}(2007){Collet}, {Asplund}, \&
  {Trampedach}}]{Collet2007}
{Collet}, R., {Asplund}, M., \& {Trampedach}, R. 2007, \aap, 469, 687

\bibitem[{{Collet} {et~al.}(2011){Collet}, {Magic}, \& {Asplund}}]{Collet2011}
{Collet}, R., {Magic}, Z., \& {Asplund}, M. 2011, in Journal of Physics
  Conference Series, Vol. 328, Journal of Physics Conference Series, 012003

\bibitem[{{Courant} {et~al.}(1967){Courant}, {Friedrichs}, \&
  {Lewy}}]{Courant1967}
{Courant}, R., {Friedrichs}, K., \& {Lewy}, H. 1967, IBM Journal of Research
  and Development, 11, 215

\bibitem[{{Dappen} {et~al.}(1987){Dappen}, {Anderson}, \&
  {Mihalas}}]{Dappen1987}
{Dappen}, W., {Anderson}, L., \& {Mihalas}, D. 1987, \apj, 319, 195

\bibitem[{{de Jong} {et~al.}(2019){de Jong}, {Agertz}, {Berbel}, {Aird},
  {Alexander}, {Amarsi}, {Anders}, {Andrae}, {Ansarinejad}, {Ansorge},
  {Antilogus}, {Anwand-Heerwart}, {Arentsen}, {Arnadottir}, {Asplund}, {Auger},
  {Azais}, {Baade}, {Baker}, {Baker}, {Balbinot}, {Baldry}, {Banerji},
  {Barden}, {Barklem}, {Barth{\'e}l{\'e}my-Mazot}, {Battistini}, {Bauer},
  {Bell}, {Bellido-Tirado}, {Bellstedt}, {Belokurov}, {Bensby}, {Bergemann},
  {Bestenlehner}, {Bielby}, {Bilicki}, {Blake}, {Bland-Hawthorn}, {Boeche},
  {Boland}, {Boller}, {Bongard}, {Bongiorno}, {Bonifacio}, {Boudon}, {Brooks},
  {Brown}, {Brown}, {Br{\"u}ggen}, {Brynnel}, {Brzeski}, {Buchert},
  {Buschkamp}, {Caffau}, {Caillier}, {Carrick}, {Casagrande}, {Case}, {Casey},
  {Cesarini}, {Cescutti}, {Chapuis}, {Chiappini}, {Childress}, {Christlieb},
  {Church}, {Cioni}, {Cluver}, {Colless}, {Collett}, {Comparat}, {Cooper},
  {Couch}, {Courbin}, {Croom}, {Croton}, {Daguis{\'e}}, {Dalton}, {Davies},
  {Davis}, {de Laverny}, {Deason}, {Dionies}, {Disseau}, {Doel}, {D{\"o}scher},
  {Driver}, {Dwelly}, {Eckert}, {Edge}, {Edvardsson}, {Youssoufi}, {Elhaddad},
  {Enke}, {Erfanianfar}, {Farrell}, {Fechner}, {Feiz}, {Feltzing}, {Ferreras},
  {Feuerstein}, {Feuillet}, {Finoguenov}, {Ford}, {Fotopoulou}, {Fouesneau},
  {Frenk}, {Frey}, {Gaessler}, {Geier}, {Gentile Fusillo}, {Gerhard},
  {Giannantonio}, {Giannone}, {Gibson}, {Gillingham},
  {Gonz{\'a}lez-Fern{\'a}ndez}, {Gonzalez-Solares}, {Gottloeber}, {Gould},
  {Grebel}, {Gueguen}, {Guiglion}, {Haehnelt}, {Hahn}, {Hansen}, {Hartman},
  {Hauptner}, {Hawkins}, {Haynes}, {Haynes}, {Heiter}, {Helmi}, {Aguayo},
  {Hewett}, {Hinton}, {Hobbs}, {Hoenig}, {Hofman}, {Hook}, {Hopgood},
  {Hopkins}, {Hourihane}, {Howes}, {Howlett}, {Huet}, {Irwin}, {Iwert},
  {Jablonka}, {Jahn}, {Jahnke}, {Jarno}, {Jin}, {Jofre}, {Johl}, {Jones},
  {J{\"o}nsson}, {Jordan}, {Karovicova}, {Khalatyan}, {Kelz}, {Kennicutt},
  {King}, {Kitaura}, {Klar}, {Klauser}, {Kneib}, {Koch}, {Koposov},
  {Kordopatis}, {Korn}, {Kosmalski}, {Kotak}, {Kovalev}, {Kreckel}, {Kripak},
  {Krumpe}, {Kuijken}, {Kunder}, {Kushniruk}, {Lam}, {Lamer}, {Laurent},
  {Lawrence}, {Lehmitz}, {Lemasle}, {Lewis}, {Li}, {Lidman}, {Lind}, {Liske},
  {Lizon}, {Loveday}, {Ludwig}, {McDermid}, {Maguire}, {Mainieri}, {Mali},
  {Mandel}, {Mandel}, {Mannering}, {Martell}, {Martinez Delgado}, {Matijevic},
  {McGregor}, {McMahon}, {McMillan}, {Mena}, {Merloni}, {Meyer}, {Michel},
  {Micheva}, {Migniau}, {Minchev}, {Monari}, {Muller}, {Murphy},
  {Muthukrishna}, {Nandra}, {Navarro}, {Ness}, {Nichani}, {Nichol}, {Nicklas},
  {Niederhofer}, {Norberg}, {Obreschkow}, {Oliver}, {Owers}, {Pai},
  {Pankratow}, {Parkinson}, {Paschke}, {Paterson}, {Pecontal}, {Parry},
  {Phillips}, {Pillepich}, {Pinard}, {Pirard}, {Piskunov}, {Plank},
  {Pl{\"u}schke}, {Pons}, {Popesso}, {Power}, {Pragt}, {Pramskiy}, {Pryer},
  {Quattri}, {Queiroz}, {Quirrenbach}, {Rahurkar}, {Raichoor}, {Ramstedt},
  {Rau}, {Recio-Blanco}, {Reiss}, {Renaud}, {Revaz}, {Rhode}, {Richard},
  {Richter}, {Rix}, {Robotham}, {Roelfsema}, {Romaniello}, {Rosario},
  {Rothmaier}, {Roukema}, {Ruchti}, {Rupprecht}, {Rybizki}, {Ryde}, {Saar},
  {Sadler}, {Sahl{\'e}n}, {Salvato}, {Sassolas}, {Saunders}, {Saviauk},
  {Sbordone}, {Schmidt}, {Schnurr}, {Scholz}, {Schwope}, {Seifert}, {Shanks},
  {Sheinis}, {Sivov}, {Sk{\'u}lad{\'o}ttir}, {Smartt}, {Smedley}, {Smith},
  {Smith}, {Sorce}, {Spitler}, {Starkenburg}, {Steinmetz}, {Stilz}, {Storm},
  {Sullivan}, {Sutherland}, {Swann}, {Tamone}, {Taylor}, {Teillon}, {Tempel},
  {ter Horst}, {Thi}, {Tolstoy}, {Trager}, {Traven}, {Tremblay}, {Tresse},
  {Valentini}, {van de Weygaert}, {van den Ancker}, {Veljanoski}, {Venkatesan},
  {Wagner}, {Wagner}, {Walcher}, {Waller}, {Walton}, {Wang}, {Winkler},
  {Wisotzki}, {Worley}, {Worseck}, {Xiang}, {Xu}, {Yong}, {Zhao}, {Zheng},
  {Zscheyge}, \& {Zucker}}]{deJong2019}
{de Jong}, R.~S., {Agertz}, O., {Berbel}, A.~A., {et~al.} 2019, The Messenger,
  175, 3

\bibitem[{{Doyle}(1968)}]{Doyle1968}
{Doyle}, R. 1968, \jqsrt, 8, 1555

\bibitem[{{Dullemond} \& {Springel}(2011)}]{Dullemond2011}
{Dullemond}, C.~P. \& {Springel}, V. 2011, Lecture Numerical Fluid Dynamics

\bibitem[{{Freytag} {et~al.}(2012){Freytag}, {Steffen}, {Ludwig},
  {Wedemeyer-B{\"o}hm}, {Schaffenberger}, \& {Steiner}}]{Freytag2012}
{Freytag}, B., {Steffen}, M., {Ludwig}, H.~G., {et~al.} 2012, Journal of
  Computational Physics, 231, 919

\bibitem[{{Fromang} {et~al.}(2006){Fromang}, {Hennebelle}, \&
  {Teyssier}}]{Fromang2006}
{Fromang}, S., {Hennebelle}, P., \& {Teyssier}, R. 2006, \aap, 457, 371

\bibitem[{{Fuhrmann} {et~al.}(1993){Fuhrmann}, {Axer}, \&
  {Gehren}}]{Fuhrmann1993}
{Fuhrmann}, K., {Axer}, M., \& {Gehren}, T. 1993, \aap, 271, 451

\bibitem[{{Gallagher} {et~al.}(2020){Gallagher}, {Bergemann}, {Collet}, {Plez},
  {Leenaarts}, {Carlsson}, {Yakovleva}, \& {Belyaev}}]{Gallagher2020}
{Gallagher}, A.~J., {Bergemann}, M., {Collet}, R., {et~al.} 2020, \aap, 634,
  A55

\bibitem[{{Gent} {et~al.}(2022){Gent}, {Bergemann}, {Serenelli}, {Casagrande},
  {Gerber}, {Heiter}, {Kovalev}, {Morel}, {Nardetto}, {Adibekyan}, {Silva
  Aguirre}, {Asplund}, {Belkacem}, {del Burgo}, {Bigot}, {Chiavassa},
  {Rodr{\'\i}guez D{\'\i}az}, {Goupil}, {Gonz{\'a}lez Hern{\'a}ndez},
  {Mourard}, {Merle}, {M{\'e}sz{\'a}ros}, {Marshall}, {Ouazzani}, {Plez},
  {Reese}, {Trampedach}, \& {Tsantaki}}]{Gent2022}
{Gent}, M.~R., {Bergemann}, M., {Serenelli}, A., {et~al.} 2022, \aap, 658, A147

\bibitem[{{Gerber} {et~al.}(2023){Gerber}, {Magg}, {Plez}, {Bergemann},
  {Heiter}, {Olander}, \& {Hoppe}}]{Gerber2023}
{Gerber}, J.~M., {Magg}, E., {Plez}, B., {et~al.} 2023, \aap, 669, A43

\bibitem[{{Gray}(1992)}]{Gray1992}
{Gray}, D.~F. 1992, {The observation and analysis of stellar photospheres.},
  Vol.~20

\bibitem[{{Grevesse} \& {Sauval}(1998)}]{Grevesse1998}
{Grevesse}, N. \& {Sauval}, A.~J. 1998, \ssr, 85, 161

\bibitem[{{Grupp}(2004)}]{Grupp2004}
{Grupp}, F. 2004, \aap, 420, 289

\bibitem[{{Gudiksen} {et~al.}(2011){Gudiksen}, {Carlsson}, {Hansteen}, {Hayek},
  {Leenaarts}, \& {Mart{\'\i}nez-Sykora}}]{Gudiksen2011}
{Gudiksen}, B.~V., {Carlsson}, M., {Hansteen}, V.~H., {et~al.} 2011, \aap, 531,
  A154

\bibitem[{{Gustafsson} {et~al.}(1975){Gustafsson}, {Bell}, {Eriksson}, \&
  {Nordlund}}]{Gustafsson1975}
{Gustafsson}, B., {Bell}, R.~A., {Eriksson}, K., \& {Nordlund}, A. 1975, \aap,
  42, 407

\bibitem[{{Gustafsson} {et~al.}(2008){Gustafsson}, {Edvardsson}, {Eriksson},
  {J{\o}rgensen}, {Nordlund}, \& {Plez}}]{Gustafsson2008}
{Gustafsson}, B., {Edvardsson}, B., {Eriksson}, K., {et~al.} 2008, \aap, 486,
  951

\bibitem[{{Haberreiter} {et~al.}(2008){Haberreiter}, {Schmutz}, \&
  {Hubeny}}]{Haberreiter2008}
{Haberreiter}, M., {Schmutz}, W., \& {Hubeny}, I. 2008, \aap, 492, 833

\bibitem[{Hartigan \& Wong(1979)}]{Hartigan1979}
Hartigan, J.~A. \& Wong, M.~A. 1979, Journal of the Royal Statistical Society.
  Series C (Applied Statistics), 28, 100

\bibitem[{{Heiter} {et~al.}(2021){Heiter}, {Lind}, {Bergemann}, {Asplund},
  {Mikolaitis}, {Barklem}, {Masseron}, {de Laverny}, {Magrini}, {Edvardsson},
  {J{\"o}nsson}, {Pickering}, {Ryde}, {Bayo Ar{\'a}n}, {Bensby}, {Casey},
  {Feltzing}, {Jofr{\'e}}, {Korn}, {Pancino}, {Damiani}, {Lanzafame}, {Lardo},
  {Monaco}, {Morbidelli}, {Smiljanic}, {Worley}, {Zaggia}, {Randich}, \&
  {Gilmore}}]{Heiter2021}
{Heiter}, U., {Lind}, K., {Bergemann}, M., {et~al.} 2021, \aap, 645, A106

\bibitem[{{Henyey} {et~al.}(1965){Henyey}, {Vardya}, \&
  {Bodenheimer}}]{Henyey1965}
{Henyey}, L., {Vardya}, M.~S., \& {Bodenheimer}, P. 1965, \apj, 142, 841

\bibitem[{{Houdek} {et~al.}(2017){Houdek}, {Trampedach}, {Aarslev}, \&
  {Christensen-Dalsgaard}}]{Houdek2017}
{Houdek}, G., {Trampedach}, R., {Aarslev}, M.~J., \& {Christensen-Dalsgaard},
  J. 2017, \mnras, 464, L124

\bibitem[{{Irwin}(1981)}]{Irwin1981}
{Irwin}, A.~W. 1981, \apjs, 45, 621

\bibitem[{{John}(1975{\natexlab{a}})}]{John1975b}
{John}, T.~L. 1975{\natexlab{a}}, \mnras, 172, 305

\bibitem[{{John}(1975{\natexlab{b}})}]{John1975a}
{John}, T.~L. 1975{\natexlab{b}}, \mnras, 170, 5

\bibitem[{{John}(1994)}]{John1994}
{John}, T.~L. 1994, \mnras, 269, 871

\bibitem[{{Karzas} \& {Latter}(1961)}]{Karzas1961}
{Karzas}, W.~J. \& {Latter}, R. 1961, \apjs, 6, 167

\bibitem[{{Kollmeier} {et~al.}(2017){Kollmeier}, {Zasowski}, {Rix}, {Johns},
  {Anderson}, {Drory}, {Johnson}, {Pogge}, {Bird}, {Blanc}, {Brownstein},
  {Crane}, {De Lee}, {Klaene}, {Kreckel}, {MacDonald}, {Merloni}, {Ness},
  {O'Brien}, {Sanchez-Gallego}, {Sayres}, {Shen}, {Thakar}, {Tkachenko},
  {Aerts}, {Blanton}, {Eisenstein}, {Holtzman}, {Maoz}, {Nandra}, {Rockosi},
  {Weinberg}, {Bovy}, {Casey}, {Chaname}, {Clerc}, {Conroy}, {Eracleous},
  {G{\"a}nsicke}, {Hekker}, {Horne}, {Kauffmann}, {McQuinn}, {Pellegrini},
  {Schinnerer}, {Schlafly}, {Schwope}, {Seibert}, {Teske}, \& {van
  Saders}}]{Kollmeier2017}
{Kollmeier}, J.~A., {Zasowski}, G., {Rix}, H.-W., {et~al.} 2017, arXiv
  e-prints, arXiv:1711.03234

\bibitem[{{Kurucz}(1979)}]{Kurucz1979}
{Kurucz}, R.~L. 1979, \apjs, 40, 1

\bibitem[{{Kurucz} {et~al.}(1984){Kurucz}, {Furenlid}, {Brault}, \&
  {Testerman}}]{Kurucz1984}
{Kurucz}, R.~L., {Furenlid}, I., {Brault}, J., \& {Testerman}, L. 1984, {Solar
  flux atlas from 296 to 1300 nm} (National Solar Observatory Atlas, Sunspot,
  New Mexico: National Solar Observatory, 1984)

\bibitem[{{Kurucz} {et~al.}(1987){Kurucz}, {van Dishoeck}, \&
  {Tarafdar}}]{Kurucz1987}
{Kurucz}, R.~L., {van Dishoeck}, E.~F., \& {Tarafdar}, S.~P. 1987, \apj, 322,
  992

\bibitem[{{Lagae} {et~al.}(2023){Lagae}, {Amarsi}, {Rodr{\'\i}guez D{\'\i}az},
  {Lind}, {Nordlander}, {Hansen}, \& {Heger}}]{Lagae2023}
{Lagae}, C., {Amarsi}, A.~M., {Rodr{\'\i}guez D{\'\i}az}, L.~F., {et~al.} 2023,
  \aap, 672, A90

\bibitem[{{Leenaarts} \& {Carlsson}(2009)}]{Leenaarts2009}
{Leenaarts}, J. \& {Carlsson}, M. 2009, in Astronomical Society of the Pacific
  Conference Series, Vol. 415, The Second Hinode Science Meeting: Beyond
  Discovery-Toward Understanding, ed. B.~{Lites}, M.~{Cheung}, T.~{Magara},
  J.~{Mariska}, \& K.~{Reeves}, 87

\bibitem[{{LeVeque}(1997)}]{LeVeque1997}
{LeVeque}, R.~J. 1997, Journal of Computational Physics, 131, 327

\bibitem[{{Lind} \& {Amarsi}(2024)}]{Lind2024}
{Lind}, K. \& {Amarsi}, A.~M. 2024, arXiv e-prints, arXiv:2401.00697

\bibitem[{{Ludwig}(1992)}]{Ludwig1992}
{Ludwig}, H.-G. 1992, PhD thesis, Kiel University

\bibitem[{{Ludwig} {et~al.}(2009{\natexlab{a}}){Ludwig}, {Behara}, {Steffen},
  \& {Bonifacio}}]{Ludwig2009b}
{Ludwig}, H.~G., {Behara}, N.~T., {Steffen}, M., \& {Bonifacio}, P.
  2009{\natexlab{a}}, \aap, 502, L1

\bibitem[{{Ludwig} {et~al.}(2009{\natexlab{b}}){Ludwig}, {Caffau}, {Steffen},
  {Freytag}, {Bonifacio}, \& {Ku{\v{c}}inskas}}]{Ludwig2009}
{Ludwig}, H.~G., {Caffau}, E., {Steffen}, M., {et~al.} 2009{\natexlab{b}},
  \memsai, 80, 711

\bibitem[{{Ludwig} \& {Steffen}(2016)}]{Ludwig2016}
{Ludwig}, H.~G. \& {Steffen}, M. 2016, Astronomische Nachrichten, 337, 844

\bibitem[{{Ludwig} {et~al.}(2023){Ludwig}, {Steffen}, \&
  {Freytag}}]{Ludwig2023}
{Ludwig}, H.~G., {Steffen}, M., \& {Freytag}, B. 2023, \aap, 679, A65

\bibitem[{{Magg} {et~al.}(2022){Magg}, {Bergemann}, {Serenelli}, {Bautista},
  {Plez}, {Heiter}, {Gerber}, {Ludwig}, {Basu}, {Ferguson}, {Gallego},
  {Gamrath}, {Palmeri}, \& {Quinet}}]{Magg2022}
{Magg}, E., {Bergemann}, M., {Serenelli}, A., {et~al.} 2022, \aap, 661, A140

\bibitem[{{Magic} {et~al.}(2013){Magic}, {Collet}, {Asplund}, {Trampedach},
  {Hayek}, {Chiavassa}, {Stein}, \& {Nordlund}}]{Magic2013a}
{Magic}, Z., {Collet}, R., {Asplund}, M., {et~al.} 2013, \aap, 557, A26

\bibitem[{{Mashonkina} {et~al.}(2008){Mashonkina}, {Zhao}, {Gehren}, {Aoki},
  {Bergemann}, {Noguchi}, {Shi}, {Takada-Hidai}, \& {Zhang}}]{Mashonkina2008}
{Mashonkina}, L., {Zhao}, G., {Gehren}, T., {et~al.} 2008, \aap, 478, 529

\bibitem[{{McLaughlin} {et~al.}(2017){McLaughlin}, {Stancil}, {Sadeghpour}, \&
  {Forrey}}]{McLaughlin2017}
{McLaughlin}, B.~M., {Stancil}, P.~C., {Sadeghpour}, H.~R., \& {Forrey}, R.~C.
  2017, Journal of Physics B Atomic Molecular Physics, 50, 114001

\bibitem[{{M{\'e}sz{\'a}ros} {et~al.}(2012){M{\'e}sz{\'a}ros}, {Allende
  Prieto}, {Edvardsson}, {Castelli}, {Garc{\'\i}a P{\'e}rez}, {Gustafsson},
  {Majewski}, {Plez}, {Schiavon}, {Shetrone}, \& {de Vicente}}]{Meszaros2012}
{M{\'e}sz{\'a}ros}, S., {Allende Prieto}, C., {Edvardsson}, B., {et~al.} 2012,
  \aj, 144, 120

\bibitem[{{Mihalas}(1965)}]{Mihalas1965}
{Mihalas}, D. 1965, \apjs, 9, 321

\bibitem[{{Mihalas} {et~al.}(1988){Mihalas}, {Dappen}, \&
  {Hummer}}]{Mihalas1988}
{Mihalas}, D., {Dappen}, W., \& {Hummer}, D.~G. 1988, \apj, 331, 815

\bibitem[{{Mosumgaard} {et~al.}(2020){Mosumgaard}, {J{\o}rgensen}, {Weiss},
  {Silva Aguirre}, \& {Christensen-Dalsgaard}}]{Mosumgaard2020}
{Mosumgaard}, J.~R., {J{\o}rgensen}, A. C.~S., {Weiss}, A., {Silva Aguirre},
  V., \& {Christensen-Dalsgaard}, J. 2020, \mnras, 491, 1160

\bibitem[{{Neckel} \& {Labs}(1994)}]{Neckel1994}
{Neckel}, H. \& {Labs}, D. 1994, \solphys, 153, 91

\bibitem[{{Nordlander} {et~al.}(2017){Nordlander}, {Amarsi}, {Lind}, {Asplund},
  {Barklem}, {Casey}, {Collet}, \& {Leenaarts}}]{Nordlander2017}
{Nordlander}, T., {Amarsi}, A.~M., {Lind}, K., {et~al.} 2017, \aap, 597, A6

\bibitem[{{Nordlund}(1982)}]{Nordlund1982}
{Nordlund}, A. 1982, \aap, 107, 1

\bibitem[{{Nordlund} \& {Galsgaard}(1995)}]{Nordlund1995}
{Nordlund}, {\AA}. \& {Galsgaard}, K. 1995, {A 3D MHD code for Parallel
  Computers}, Tech. rep., {Niels Bohr Institute, University of Copenhagen}

\bibitem[{{Nordlund} {et~al.}(2018){Nordlund}, {Ramsey}, {Popovas}, \&
  {K{\"u}ffmeier}}]{Nordlund2018}
{Nordlund}, {\r{A}}., {Ramsey}, J.~P., {Popovas}, A., \& {K{\"u}ffmeier}, M.
  2018, \mnras, 477, 624

\bibitem[{{Nordlund} {et~al.}(2009){Nordlund}, {Stein}, \&
  {Asplund}}]{Nordlund2009}
{Nordlund}, {\r{A}}., {Stein}, R.~F., \& {Asplund}, M. 2009, Living Reviews in
  Solar Physics, 6, 2

\bibitem[{{Peach}(1970)}]{Peach1970}
{Peach}, G. 1970, \memras, 73, 1

\bibitem[{{Pereira} {et~al.}(2013){Pereira}, {Asplund}, {Collet}, {Thaler},
  {Trampedach}, \& {Leenaarts}}]{Pereira2013}
{Pereira}, T.~M.~D., {Asplund}, M., {Collet}, R., {et~al.} 2013, \aap, 554,
  A118

\bibitem[{{Plez}(2012)}]{Plez2012}
{Plez}, B. 2012, {Turbospectrum: Code for spectral synthesis}, Astrophysics
  Source Code Library, record ascl:1205.004

\bibitem[{{Rauer} {et~al.}(2014){Rauer}, {Catala}, {Aerts}, {Appourchaux},
  {Benz}, {Brandeker}, {Christensen-Dalsgaard}, {Deleuil}, {Gizon}, {Goupil},
  {G{\"u}del}, {Janot-Pacheco}, {Mas-Hesse}, {Pagano}, {Piotto}, {Pollacco},
  {Santos}, {Smith}, {Su{\'a}rez}, {Szab{\'o}}, {Udry}, {Adibekyan}, {Alibert},
  {Almenara}, {Amaro-Seoane}, {Eiff}, {Asplund}, {Antonello}, {Barnes},
  {Baudin}, {Belkacem}, {Bergemann}, {Bihain}, {Birch}, {Bonfils}, {Boisse},
  {Bonomo}, {Borsa}, {Brand{\~a}o}, {Brocato}, {Brun}, {Burleigh}, {Burston},
  {Cabrera}, {Cassisi}, {Chaplin}, {Charpinet}, {Chiappini}, {Church},
  {Csizmadia}, {Cunha}, {Damasso}, {Davies}, {Deeg}, {D{\'\i}az}, {Dreizler},
  {Dreyer}, {Eggenberger}, {Ehrenreich}, {Eigm{\"u}ller}, {Erikson}, {Farmer},
  {Feltzing}, {de Oliveira Fialho}, {Figueira}, {Forveille}, {Fridlund},
  {Garc{\'\i}a}, {Giommi}, {Giuffrida}, {Godolt}, {Gomes da Silva}, {Granzer},
  {Grenfell}, {Grotsch-Noels}, {G{\"u}nther}, {Haswell}, {Hatzes},
  {H{\'e}brard}, {Hekker}, {Helled}, {Heng}, {Jenkins}, {Johansen},
  {Khodachenko}, {Kislyakova}, {Kley}, {Kolb}, {Krivova}, {Kupka}, {Lammer},
  {Lanza}, {Lebreton}, {Magrin}, {Marcos-Arenal}, {Marrese}, {Marques},
  {Martins}, {Mathis}, {Mathur}, {Messina}, {Miglio}, {Montalban}, {Montalto},
  {Monteiro}, {Moradi}, {Moravveji}, {Mordasini}, {Morel}, {Mortier},
  {Nascimbeni}, {Nelson}, {Nielsen}, {Noack}, {Norton}, {Ofir}, {Oshagh},
  {Ouazzani}, {P{\'a}pics}, {Parro}, {Petit}, {Plez}, {Poretti}, {Quirrenbach},
  {Ragazzoni}, {Raimondo}, {Rainer}, {Reese}, {Redmer}, {Reffert},
  {Rojas-Ayala}, {Roxburgh}, {Salmon}, {Santerne}, {Schneider}, {Schou},
  {Schuh}, {Schunker}, {Silva-Valio}, {Silvotti}, {Skillen}, {Snellen}, {Sohl},
  {Sousa}, {Sozzetti}, {Stello}, {Strassmeier}, {{\v{S}}vanda}, {Szab{\'o}},
  {Tkachenko}, {Valencia}, {Van Grootel}, {Vauclair}, {Ventura}, {Wagner},
  {Walton}, {Weingrill}, {Werner}, {Wheatley}, \& {Zwintz}}]{Rauer2014}
{Rauer}, H., {Catala}, C., {Aerts}, C., {et~al.} 2014, Experimental Astronomy,
  38, 249

\bibitem[{{Roe}(1986)}]{Roe1986}
{Roe}, P.~L. 1986, Annual Review of Fluid Mechanics, 18, 337

\bibitem[{{Rosenthal} {et~al.}(1999){Rosenthal}, {Christensen-Dalsgaard},
  {Nordlund}, {Stein}, \& {Trampedach}}]{Rosenthal1999}
{Rosenthal}, C.~S., {Christensen-Dalsgaard}, J., {Nordlund}, {\r{A}}., {Stein},
  R.~F., \& {Trampedach}, R. 1999, \aap, 351, 689

\bibitem[{{Ruchti} {et~al.}(2013){Ruchti}, {Bergemann}, {Serenelli},
  {Casagrande}, \& {Lind}}]{Ruchti2013}
{Ruchti}, G.~R., {Bergemann}, M., {Serenelli}, A., {Casagrande}, L., \& {Lind},
  K. 2013, \mnras, 429, 126

\bibitem[{{Rybicki} \& {Hummer}(1992)}]{Rybicki1992}
{Rybicki}, G.~B. \& {Hummer}, D.~G. 1992, \aap, 262, 209

\bibitem[{{Samadi} {et~al.}(2003){Samadi}, {Nordlund}, {Stein}, {Goupil}, \&
  {Roxburgh}}]{Samadi2003}
{Samadi}, R., {Nordlund}, {\r{A}}., {Stein}, R.~F., {Goupil}, M.~J., \&
  {Roxburgh}, I. 2003, \aap, 404, 1129

\bibitem[{{Sbordone} {et~al.}(2010){Sbordone}, {Bonifacio}, {Caffau}, {Ludwig},
  {Behara}, {Gonz{\'a}lez Hern{\'a}ndez}, {Steffen}, {Cayrel}, {Freytag},
  {van't Veer}, {Molaro}, {Plez}, {Sivarani}, {Spite}, {Spite}, {Beers},
  {Christlieb}, {Fran{\c{c}}ois}, \& {Hill}}]{Sbordone2010}
{Sbordone}, L., {Bonifacio}, P., {Caffau}, E., {et~al.} 2010, \aap, 522, A26

\bibitem[{{Sch{\"o}nrich} \& {Bergemann}(2014)}]{Schoenrich2014}
{Sch{\"o}nrich}, R. \& {Bergemann}, M. 2014, \mnras, 443, 698

\bibitem[{{Schwarzschild}(1958)}]{Schwarzschild1958}
{Schwarzschild}, M. 1958, {Structure and evolution of the stars.}

\bibitem[{{Seaton} {et~al.}(1994){Seaton}, {Yan}, {Mihalas}, \&
  {Pradhan}}]{Seaton1994}
{Seaton}, M.~J., {Yan}, Y., {Mihalas}, D., \& {Pradhan}, A.~K. 1994, \mnras,
  266, 805

\bibitem[{{Short} \& {Hauschildt}(2005)}]{Short2005}
{Short}, C.~I. \& {Hauschildt}, P.~H. 2005, \apj, 618, 926

\bibitem[{{Stein} \& {Nordlund}(1998)}]{Stein1998}
{Stein}, R.~F. \& {Nordlund}, {\r{A}}. 1998, \apj, 499, 914

\bibitem[{{Stein} \& {Nordlund}(2001)}]{Stein2001}
{Stein}, R.~F. \& {Nordlund}, {\r{A}}. 2001, \apj, 546, 585

\bibitem[{{Straumit} {et~al.}(2022){Straumit}, {Tkachenko}, {Gebruers},
  {Audenaert}, {Xiang}, {Zari}, {Aerts}, {Johnson}, {Kollmeier}, {Rix},
  {Beaton}, {Van Saders}, {Teske}, {Roman-Lopes}, {Ting}, \&
  {Rom{\'a}n-Z{\'u}{\~n}iga}}]{Straumit2022}
{Straumit}, I., {Tkachenko}, A., {Gebruers}, S., {et~al.} 2022, \aj, 163, 236

\bibitem[{{van Leer}(1977)}]{vanLeer1977}
{van Leer}, B. 1977, Journal of Computational Physics, 23, 276

\bibitem[{{V{\"o}gler} {et~al.}(2004){V{\"o}gler}, {Bruls}, \&
  {Sch{\"u}ssler}}]{Vogler2004}
{V{\"o}gler}, A., {Bruls}, J.~H.~M.~J., \& {Sch{\"u}ssler}, M. 2004, \aap, 421,
  741

\bibitem[{{V{\"o}gler} {et~al.}(2005){V{\"o}gler}, {Shelyag}, {Sch{\"u}ssler},
  {Cattaneo}, {Emonet}, \& {Linde}}]{Vogler2005}
{V{\"o}gler}, A., {Shelyag}, S., {Sch{\"u}ssler}, M., {et~al.} 2005, \aap, 429,
  335

\bibitem[{{Williamson}(1980)}]{Williamson1980}
{Williamson}, J.~H. 1980, Journal of Computational Physics, 35, 48

\bibitem[{{Wolf}(1983)}]{Wolf1983}
{Wolf}, B.~E. 1983, \aap, 127, 93

\bibitem[{{Young} \& {Short}(2014)}]{Young2014}
{Young}, M.~E. \& {Short}, C.~I. 2014, \apj, 787, 43

\end{thebibliography}
\begin{appendix}
\section{Supplementary material}
\begin{table*}
\centering                          
\setlength{\tabcolsep}{3.0pt}
\caption[]{Model properties of \texttt{M3DIS} solar models.}
\label{tab:stellar_parameters}
\begin{tabular}{l c c c c c c c  c}            
\hline\hline                        
\smallskip
Model & $\rm T_{eff}$ & $\rm log(g)$ & $\rm [Fe/H]$ & composition & HD solver & EoS &  $\rm N_{bins}$ \\
\hline                              
 \texttt{M3DIS} (A) & $\rm 5923.5 \pm  7.7\ K$ & 4.44 & 0.0 & \cite{Magg2022} & HLLC & \cite{Gerber2023}* & 7 \\
 \texttt{M3DIS} (B) & $\rm 5858.9 \pm 14.8\ K$ & 4.44 & 0.0 & \cite{Magg2022} & HLLC & \cite{Gerber2023}* & 7 \\
 \texttt{M3DIS} (C) & $\rm 5786.0 \pm  9.6\ K$ & 4.44 & 0.0 & \cite{Magg2022} & HLLC & \cite{Gerber2023}* & 7 \\
 \texttt{M3DIS} (D) & $\rm 5808.2 \pm 14.8\ K$ & 4.44 & 0.0 & \cite{Magg2022} & HLLC & \cite{Gerber2023}* & 7 \\

\texttt{Stagger} & $\rm 5768.51\ K$ & 4.44 & 0.0 & \cite{Asplund2009} & Runge-Kutta & \cite{Mihalas1988} & 12 \\
\texttt{CO$^{5}$BOLD} & $\rm 5781\ K$ & 4.44 & 0.0 & \citet{Grevesse1998}, except CNO & Roe Riemann  & \cite{Wolf1983} & 12 \\

\hline
\end{tabular} \\
\tablefoot{
    \tablefoottext{*}{The EoS is computed using \texttt{Turbospectrum2020}, which is consistent with \texttt{MARCS} in terms of chemical equilibrium.}
    
    For all models presented in this table the opacities from \cite{Gustafsson2008} were used.
} 
\end{table*}

\begin{table}
\begin{minipage}{\linewidth}
\renewcommand{\footnoterule}{} 
\setlength{\tabcolsep}{3pt}
\caption{Diagnostic lines of Fe I adopted for tests presented in this work.}
\label{tab:lines}     
\begin{center}
\begin{tabular}{l ccc}
\noalign{\smallskip}\hline\noalign{\smallskip}  
$\lambda$ [\AA] & state (low) & state (up)&  $\log gf$ \\
\noalign{\smallskip}\hline\noalign{\smallskip}
     5242.491 &    a1I6  &  z1H5* & -0.830  \\     
     5379.574 &    b1G4  &  z1H5* & -1.514  \\    
     5618.632 &   z3P2*  &   e3D2 & -1.250  \\    
     6027.051 &    c3F4  &  v3G5* & -1.089  \\    
     6065.482 &    b3F2  &  y3F2* & -1.529  \\    
     6165.360 &    c3F3  &  v3G4* & -1.473  \\    
     6173.334 &    a5P1  &  y5D0* & -2.880  \\    
     6200.312 &    b3F2  &  y3F3* & -2.433  \\    
     6213.429 &    a5P1  &  y5D1* & -2.481  \\    
     6246.318 &   z5P3*  &   e5D3 & -0.770  \\    
     6252.555 &    a3H6  &  z3G5* & -1.699  \\    
     6430.845 &    a5P3  &  y5D4* & -1.950  \\    
     6725.356 &   y5D4*  &   e3F4 & -2.100  \\    
     6750.151 &    a3P1  &  z3P1* & -2.618  \\    
     6810.262 &   y5P2*  &   e5P3 & -0.986  \\    
     6843.655 &   y3F4*  &   e3D3 & -0.730  \\    
     8471.743 &   x5D3*  &   g5D3 & -0.910  \\    
     8514.071 &    a5P2  &  z5P2* & -2.200  \\    
     8621.601 &    b3G5  &  z3G5* & -2.320  \\    
     8784.440 &   x5D3*  &   g5D4 & -1.140  \\    
     8905.991 &   x5F3*  &   e5P2 & -1.220  \\    
     8945.189 &   x5F4*  &   g5D3 & -0.230  \\
\noalign{\smallskip}\hline\noalign{\smallskip}
\end{tabular}
\end{center}
\end{minipage}
\end{table} 

\begin{figure}
    \includegraphics[width=\linewidth]{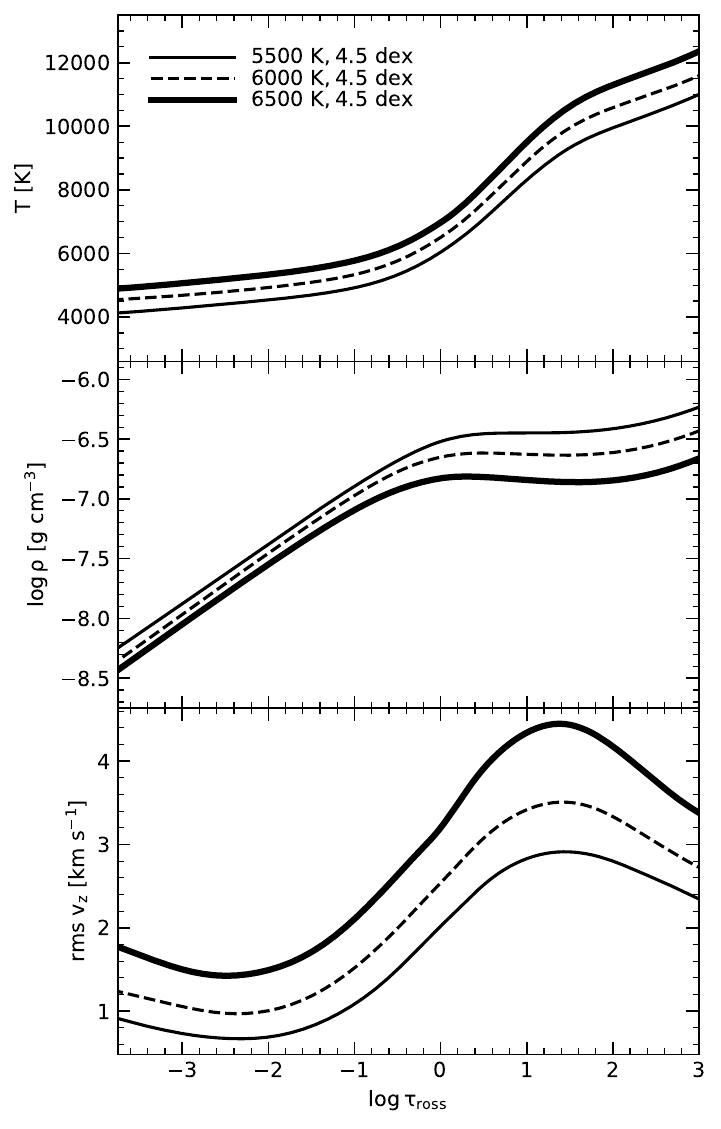}   
    \caption{Average temperature and density profiles for three selected \texttt{M3DIS} models of solar-like stars. }
\label{fig:average_profiles_others}
\end{figure}

In Fig. \ref{fig:average_profiles_others}, we show the horizontally averaged profiles of temperature (top), density (middle), and vertical velocity (bottom) against Rosseland optical depth for three models of Sun-like stars computed using our new \texttt{M3DIS} code. The profiles show that hotter model structures (corresponding to $\teff = 6500$ K) are characterised by lower densities and higher flow velocities, compared to lower $\teff$ model atmospheres. This implies that impact of convective motions and inhomogeneities on spectral lines is larger for hotter stars.  
\begin{figure}
    \includegraphics[width=\linewidth]{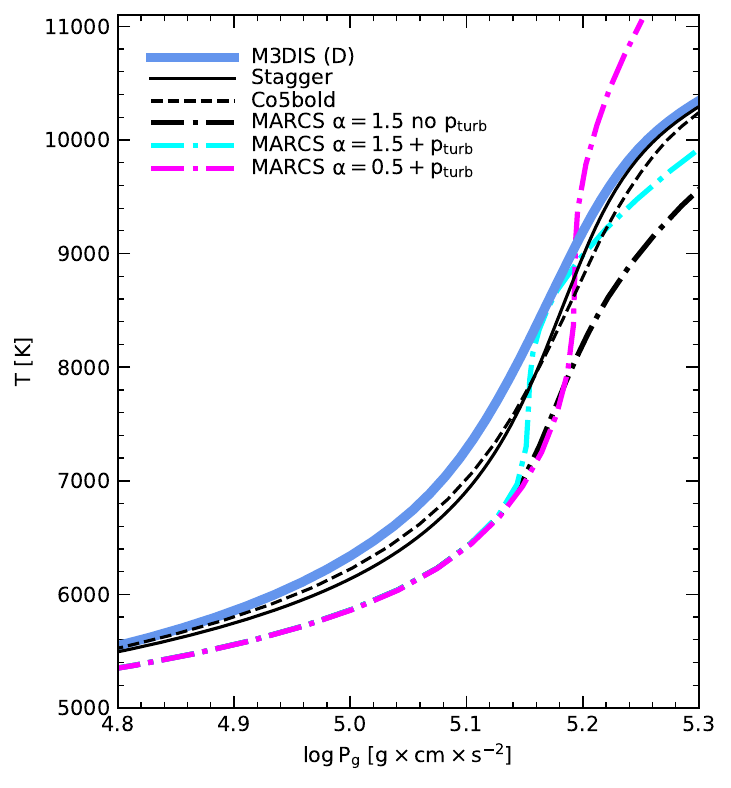}   
    \caption{Comparison of average temperature and gas pressure profiles for \texttt{M3DIS}, \texttt{STAGGER}, \texttt{CO$^5$BOLD}, and \texttt{MARCS} solar model atmospheres. We also show different versions of the \texttt{MARCS} models, computed with and without turbulent pressure, and with different mixing length parameters $\alpha$.}
\label{fig:T-Pg_3D_MARCS}
\end{figure}

In Fig. \ref{fig:T-Pg_3D_MARCS}, we compare the gas pressure versus temperature for the spatially-averaged profiles of 3D RHD solar model atmospheres computed with different codes (\texttt{M3DIS}, \texttt{STAGGER}, and \texttt{CO$^{5}$BOLD}),  with the corresponding profile of the standard 1D hydrostatic MARCS model atmosphere. The latter is characterised by systematically higher gas pressures and densities compared to the 3D models, with differences exceeding 10 $\%$ in the inner atmospheric layers at $\ltross \gtrsim 0$.

\begin{figure}
\centering
\includegraphics[width=0.91\linewidth]{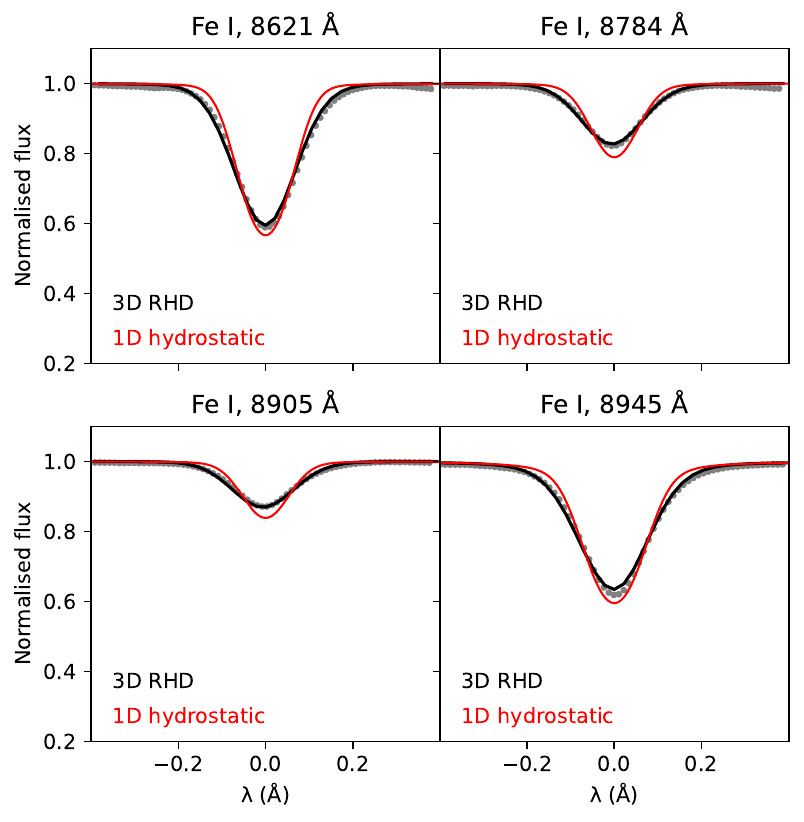}
\caption{As Fig. \ref{fig:feall}, for the remaining Fe I lines from table \ref{tab:lines}.}
\label{fig:fe_add}
\end{figure}

In Fig. \ref{fig:fe_add}, we show the 3D synthetic profiles of remaining Fe I lines from Table \ref{tab:lines} in comparison with the observed solar flux data. See Fig. \ref{fig:feall} for more details.

\end{appendix}

\end{document}